\documentclass[11pt,a4paper,oneside]{article}
\usepackage[top=3cm, bottom=3cm, left=2cm, right=2cm]{geometry}
\usepackage[english]{babel}
\usepackage[utf8]{inputenc}
\usepackage{enumerate}
\usepackage[T1]{fontenc}
\usepackage{setspace}
\usepackage{xcolor}
\usepackage{booktabs}
\usepackage{multirow}
\usepackage{xr}
\usepackage{tikz}
\usetikzlibrary{positioning}
\usetikzlibrary{fit}

\usepackage{amsthm,amsmath,amssymb,mathrsfs,dsfont,mathtools, mathabx}
\usepackage{bbm}
\usepackage{bm}
\usepackage{graphicx}
\usepackage[colorlinks=true, allcolors=blue]{hyperref}
\usepackage{comment}
\usepackage[all]{nowidow}

\usepackage[authoryear, round]{natbib}

\usepackage{authblk}
\date{ \vspace{-5mm} }
\title{Bayesian Learning of Clinically Meaningful Sepsis Phenotypes in Northern Tanzania}
\author[1]{Alexander Dombowsky}
\author[1,2]{David B. Dunson}
\author[3,4,5]{Deng B. Madut}
\author[3,4,5]{Matthew P. Rubach}
\author[1,3,6]{Amy H. Herring}
\affil[1]{{\small Department of Statistical Science, Duke University, NC, USA}}
\affil[2]{{\small Department of Mathematics, Duke University, NC, USA}}
\affil[3]{{\small Duke Global Health Institute, Duke University, NC, USA}}
\affil[4]{{\small Department of Medicine, Duke University, NC, USA}}
\affil[5]{{\small Department of Infectious Diseases \& International Health, Duke University, NC, USA}}
\affil[6]{{\small Department of Biostatistics \& Bioinformatics, Duke University, NC, USA}}

\usepackage[sf]{titlesec}
\usepackage{sectsty}
\allsectionsfont{\centering \normalfont\scshape} 

\theoremstyle{definition}

\def\bs{\boldsymbol}
\def\lb{\left\{ }
\def\rb{\right\}}
\def\R{\mathbb{R}}

\def\N{\mathcal{N}}

\def\tx{\textrm}
\def\y{\bs y}

\def\c{\bs c}
\def\s{\bs s}
\def\M{\mathcal{M}}

\def\mujl{\mu_j^{(l)}}

\def\sigmajlsq{\sigma_j^{(l)2}}
\def\xijk{\xi_j^{(k)}}
\def\taujksq{\tau_j^{(k)2}}

\begin{document}

\maketitle

\begin{abstract}
Sepsis is a life-threatening condition caused by a dysregulated host response to infection. Recently, researchers have hypothesized that sepsis consists of a heterogeneous spectrum of distinct subtypes, motivating several studies to identify clusters of sepsis patients that correspond to subtypes, with the long-term goal of using these clusters to design subtype-specific treatments. Therefore, clinicians rely on clusters having a concrete medical interpretation, usually corresponding to clinically meaningful regions of the sample space that have a concrete implication to practitioners. In this article, we propose Clustering Around Meaningful Regions (CLAMR), a Bayesian clustering approach that explicitly models the medical interpretation of each cluster center. CLAMR favors clusterings that can be summarized via meaningful feature values, leading to medically significant sepsis patient clusters. We also provide details on measuring the effect of each feature on the clustering using Bayesian hypothesis tests, so one can assess what features are relevant for cluster interpretation. Our focus is on clustering sepsis patients from Moshi, Tanzania, where patients are younger and the prevalence of HIV infection is higher than in previous sepsis subtyping cohorts.
\end{abstract}

\doublespacing

\clearpage

\section{Introduction}
For hospitals across the world, the diagnosis and rapid treatment of sepsis is of paramount importance for improving in-hospital outcomes. Broadly, sepsis is defined as organ dysfunction resulting from an immune response to an infection \citep{singer2016third} and is the cause of one-third to one-half of in-hospital deaths in the USA \citep{liu2014hospital}. The catalyst for sepsis is usually a bacterial infection \citep{bennett2019mandell}, located in the lungs, brain, urinary tract, or digestive system \citep{tintinalli2016tintinalli}, but sepsis can also be a generalized systemic infection (e.g., bloodstream infection), and microbes causing sepsis also include fungal, viral, and protozoal pathogens. The clinical manifestations are dynamic, changing over the course of the illness. If not recognized and treated early, a patient is more likely to progress to a state of circulatory collapse, known as septic shock, which significantly increases the risk of mortality \citep{singer2016third}. Sepsis is life-threatening: even minor organ dysfunction in the presence of infection has a mortality rate of at least 10\% \citep{singer2016third}, and septic shock is estimated to have as high as 25\% mortality \citep{rowan2017early}. Once detected, prompt interventions, such as antibiotics, vasopressors \citep{rhodes2017surviving}, intravenous fluids \citep{rowan2017early}, or adjunctive corticosteroids \citep{annane2018hydrocortisone} can improve outcomes. 

However, despite its prevalence and risk, therapies specifically targeted towards treatment of sepsis, including early goal-directed therapy (EGDT), have not led to improved outcomes \citep{opal2014next, rowan2017early, van2017immunopathology}. This phenomenon has been attributed to the hypothesis that the general condition of sepsis is comprised of heterogeneous and distinct subtypes
which may respond better to treatment plans specifically oriented to the pathophysiology of these respective subtypes.
For example, randomized trials of adjunctive corticosteroids in adults with sepsis had mixed results, and in pediatric patients adjunctive corticosteroids were associated with increased mortality in one subtype of septic shock \citep{wong2017pediatric}. Sepsis subtypes have been derived by several studies which clustered in-hospital cohorts and found distinct patient groups, or \textit{phenotypes}, using comparative mRNA transcriptional analysis of the patient's immune response \citep{davenport2016genomic,scicluna2017classification,wong2017pediatric,sweeney2018unsupervised}. 
Given that these descriptions of sepsis subtypes were primarily based upon immunologic characteristics, the Sepsis Endotyping in Emergency Care (SENECA) project hypothesized that subtypes could also be derived from routine clinical information that is readily extractable from patient electronic health record sources \citep{seymour2019derivation}. They described four distinct phenotypes characterized by age and renal dysfunction ($\beta$), inflammation and pulmonary dysfunction ($\gamma$), liver dysfunction and septic shock ($\delta$), and the absence of these characteristics ($\alpha$). Moreover, they found evidence that EGDT was beneficial for patients in the $\alpha$ subtype, whereas the $\delta$ subtype responded negatively to the treatment.

Our interest is in inferring phenotypes present within a patient population not adequately represented by previous studies. The Investigating Febrile Deaths in Tanazania (INDITe) study sought to describe the etiologies of fatal febrile illness in sub-Saharan Africa (sSA) with the over-arching goal of identifiying priorities for the prevention and treatment of severe infectious diseases \citep{snavely2018sociocultural}. From September 2016 to May 2019, the study collected detailed transcriptome and clinical information from $599$ febrile patients across two hospitals in Moshi, Tanzania, the administrative seat of the Kilimanjaro Region. Studies have estimated that the in-hospital mortality rate of severe sepsis in sSA is between $20$ and $60\%$ \citep{jacob2009severe,andrews2014simplified, andrews2017effect}, and the region has the highest health-related burden of sepsis in the world \citep{rudd2020global}. In contrast to the primarily European and North American populations in previous phenotype classifications such as SENECA, the INDITe cohort had markedly different immunological profiles and sepsis etiologies. For one, the average age of the INDITe cohort was 41.3, whereas for SENECA the average age was 64 \citep{seymour2019derivation}. Secondly, the prevalence of HIV infection among the INDITe participants was $38.2\%$, and of the persons living with HIV, $69.0\%$ had CD4 T-cell lymphocyte percentage indicative of advanced HIV; this prevalence of HIV and advanced HIV is much higher than the HIV prevalence in the sepsis subtyping populations from Europe and North America. In terms of etiologies, febrile patients in Kilimanjaro exhibit pathogens not common in Europe and North America, including cryptococcosis, chikungunya, and histoplasmosis \citep{crump2013etiology, rubach2015etiologies}. Due to these substantial differences, it is possible that there were endotypes present in the INDITe cohort that have not been previously cataloged. 

Clustering methods utilized by the aforementioned phenotype analyses include agglomerative hierarchical clustering \citep{jain1988algorithms}, consensus k-means \citep{wilkerson2010consensusclusterplus}, and a combined mapping of clustering algorithms \citep{sweeney2015combined}. Such methods can be categorized as being \textit{algorithmic} in using an iterative mechanism to minimize a clustering loss function or meet a clustering criterion. Algorithmic clustering methods can work well in certain cases but need the number of clusters to be pre-specified and typically lack a characterization of uncertainty in clustering. These points are addressed by \textit{model-based} methods which compute clusters using tools from statistical inference. Model-based clustering methods rely on 
\textit{mixture models} to simultaneously sort individuals into groups while inferring the probability distribution of data within each group. The Gaussian mixture model (GMM) results when one assumes that the features of individuals in each cluster follow a multivariate Gaussian distribution. A wide range of approaches are used to estimate the clustering of the data under a GMM using either frequentist or Bayesian inference, see \cite{maclachlanfinite}, \cite{fruhwirth2019handbook}, and \cite{wade2023bayesian} for an overview. 

This article will take a Bayesian approach to clustering. We observe data $\bs y = (\y_1, \dots, \y_n)$, where for each individual $i$, we have features $\y_{i} = (y_{i1}, \dots, y_{ip})$. 
Our goal is to group patients into $L$ phenotype clusters, denoted by $\bs c = (c_1, \dots, c_n)$, where $c_i \in \lb 1, \dots, L \rb$. 
To express uncertainty in $\bs c$ before observing $\y$, we assign a prior $\pi(\c)$, which is a probability distribution over the space of all clusterings. Incorporating information in the data, the prior is updated to obtain the posterior $\pi(\c \mid \y)$. Since this posterior tends to be analytically intractable, Bayesian inference typically relies on Markov chain Monte Carlo (MCMC) algorithms that produce samples $\c^{(t)}$ from $\pi(\c \mid \y)$ \citep{maclachlanfinite,wade2023bayesian}. From the samples $\c^{(t)}$, one obtains standard quantities in statistical inference, including a point estimate $\c^*$ of $\c$ \citep{binder1978bayesian, lau2007bayesian} and uncertainty quantification using a credible ball \citep{wade2018bayesian} or posterior similarity matrix (PSM), the latter of which quantifies $\Pr(c_i = c_j \mid \bs y)$. Endotypes have been derived using Bayesian clustering for illnesses such as cancer \citep{savage2010discovering, yuan2011patient, lock2013bayesian, wang2023bayesian}, Alzheimer's \citep{poulakis2022multi}, emphysema \citep{ross2016bayesian}, and type 2 diabetes \citep{udler2018type}. The Bayesian paradigm can allow a broad variety of feature types including time-to-event \citep{bair2004semi, raman2010infinite, ahmad2017towards} and longitudinal data \citep{poulakis2020fully, poulakis2022multi, lu2022bayesian}. 

A key step in Bayesian clustering is modeling the distribution of observations within each cluster. Within cluster $l$, data are assumed to be distributed as $f(\cdot ; \bs \theta^{(l)})$, where $f$ is some parametric probability kernel and $\bs \theta^{(l)}$ is the cluster-specific parameter. In a GMM, $f$ is the Gaussian kernel and $\bs \theta^{(l)} = (\bs \mu^{(l)}, \Sigma^{(l)})$ is the mean and covariance of the $l$th cluster. An often under-discussed aspect of Bayesian clustering is the prior specification for $\bs \theta^{(l)}$. The general consensus is to assume that $\pi(\bs \theta^{(l)})$ is conjugate to $f$ because $\pi(\c \mid \y)$ can then be sampled using a Gibbs sampler. For GMMs, the conjugate family is the Gaussian-inverse-Wishart prior.
It is a well-known fact that improper priors for the cluster-specific parameters lead to improper posteriors \citep{wasserman2000asymptotic}, so the cluster-specific priors can be weakly-informative on likely feature ranges for each cluster \citep{richardson1997bayesian} or data-based \citep{raftery1995hypothesis, bensmail1997inference}. Cluster-specific priors are inappropriate for our application because we expect to observe new phenotypes explained by the immunoprofiles of the INDITe cohort, and label-switching is still likely to occur even if substantial prior information on the phenotypes were available. Additionally, data-based priors forfeit the probabilistic updating of information that occurs in subjectivist Bayesian inference and can potentially lead to underestimation of uncertainty. 

To analyze the INDITe data, we propose a simple but novel prior for $\bs \theta^{(l)}$ that is motivated by phenotype inference. Essential to understanding our prior is the notion of meaningful regions (MRs), areas of the feature domain that have a clear clinical interpretation. For example, MRs of the $j$th feature could correspond to intervals in which clinicians can classify an individual as having diminished, neutral, or elevated expression of that feature. MRs are assumed to be known in advance, usually identified by investigators with expert knowledge of the application. We focus on the cluster center, or the mean, in a GMM. We specify the prior for the centers to be itself a mixture model comprised of non-overlapping and informative components which cover the MRs. This prior highly favors cluster centers located inside the meaningful regions, ultimately leading to clinically relevant clusters. For that reason, we call our approach CLustering Around Meaningful Regions (CLAMR). In Section \ref{sect:methods}, we introduce CLAMR, discuss default prior settings, show how our prior can test the influence of features on the clustering, and provide computational details. In Section \ref{sect:simulations}, CLAMR is evaluated by repeatedly sampling synthetic data. We display all results for applying CLAMR to the INDITe data in Section \ref{sect:indite} and provide interpretations of the phenotypes we uncover. Finally, we present concluding remarks, extensions, and generalizations in Section \ref{sect:discussion}.

\section{Clustering Around Meaningful Regions} \label{sect:methods}

\begin{figure}
    \centering
    \includegraphics[scale=0.5]{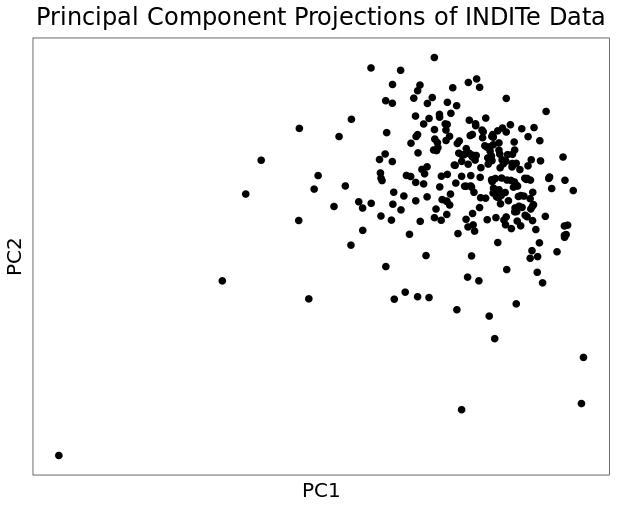}
    \caption{Projections onto the first two principal components for a subset of the numeric features from the original INDITe dataset.}
    \label{fig:indite-principal-components}
\end{figure}

\subsection{Applied Context and Motivation}

In this article, we aim to catalog sepsis phenotypes in the INDITe data which reflect the subtypes of sepsis present in northern Tanzania. The INDITe study enrolled $599$ febrile patients from two Moshi hospitals and measured both detailed RNA sequencing and routine clinical information on each participant. The clinical data are derived from two different sources. The first is a case report form (CRF) that details relevant signs and demographic information on the individual, and the second is comprehensive laboratory testing. 
From a statistical perspective, the INDITe data presents several challenges for cluster analysis. An integral factor is the lack of any clear separation in the data into isolated clusters, as can be seen in the principal component projections in Figure \ref{fig:indite-principal-components}. While there are some noticeable patterns, including a clear outlier and skewness in the density of the data, we expect there to be substantial uncertainty in cluster membership. Moreover, the sample size of the INDITe study is markedly smaller than previous phenotype analyses, particularly SENECA. For these reasons, we use Bayesian methods that can coherently incorporate uncertainty quantification to cluster the INDITe data. Furthermore, despite the lack of separation, we opt to distinguish clusters by accounting for differences in \textit{clinical interpretation}. The intuition here is that clusters should reflect a relevant medical designation that can inform treatment. Accounting for interpretations involves incorporating prior information on the features into our analysis.
Finally, an additional challenge is the selection of clinical variables that explain fundamental differences between the clusters, which can be useful for phenotype assignment of a patient. In the following sections, we discuss an approach that can account for each of these factors while leveraging the mathematical benefits of a Bayesian clustering model.

\subsection{Bayesian Probabilistic Clustering}

Let $y_{ij} \in \mathbb{R}$ denote the value of the $j$th feature for the $i$th individual, $j=1,\dots,p$ and $i=1,\dots,n$, which we collect into the matrix $\bs y = (y_{ij})_{i,j}$. We are primarily interested in inferring the phenotype clustering $\c = (c_1, \dots, c_n)$, where $c_i \in \lb 1, \dots, L \rb$ and $L > 0$, with $C_l = \lb i : c_i = l  \rb$ denoting the $l$th cluster.  
Generally, the likelihood $\pi(\y \mid \c)$ depends on some \textit{kernel family} $\mathcal{F} = \lb f(\bs \theta) : \bs \theta \in \Theta \rb$ and \textit{cluster-specific parameters} $\bs \theta = (\bs \theta^{(1)}, \dots, \bs \theta^{(L)})$, so that $\bs \theta^{(l)} \in \Theta$ for all $l$ and $(\bs y_i \mid c_i = l) \sim f(\cdot ; \bs \theta^{(l)})$. In many applications, the cluster-specific parameters are dimension dependent, so we write $\bs \theta^{(l)} = (\theta_1^{(l)}, \dots, \theta_p^{(l)})$. After specifying a prior distribution $\pi(\bs \theta)$ for the cluster-specific parameters,
the marginal posterior distribution of $\bs c$ is
\begin{equation} \label{eq:c-posterior}
    \pi(\c \mid \y) \propto \pi(\c) \prod_{l=1}^{L} \int_{\bs \theta^{(l)}} \lb  \prod_{i: c_i = l} f(\bs y_i; \bs \theta^{(l)}) \rb \pi(\bs \theta^{(l)})  d \bs \theta^{(l)}.
\end{equation} 

There are many possible choices for the clustering prior, including Dirichlet-Multinomial, the Dirichlet process \citep{ferguson1973bayesian}, and the Pitman-Yor process  \citep{pitman1997two}. 
Alternatively, when prior information on the clustering $\c$ is available, one can use informative clustering priors such as constrained clustering \citep{lu2004semi} and centered partition processes \citep{paganin2021centered}. Also relevant to the posterior of $\c$ in \eqref{eq:c-posterior} are the priors for the cluster-specific parameters, $\pi(\bs \theta^{(l)})$. One approach is to calibrate $\pi(\bs \theta^{(l)})$ to be \textit{weakly informative} for each cluster using prior information. For example, $\pi(\bs \theta^{(l)})$ can be specified to have support on a likely range of spread for cluster $l$ if such information is available \citep{richardson1997bayesian}. However, a more common approach is to assume that $\bs \theta^{(l)}$ are identically distributed for all $l$ from some \textit{base measure} $g(; \bs \nu)$ with \textit{hyperparameters} $\bs \nu$, i.e. $\pi(\bs \theta^{(l)}) = g(\bs \theta^{(l)}; \bs \nu)$ for all $l$ \citep{wade2023bayesian}. The choice of $\bs \nu$ in such models is a frequent topic of discussion in the literature. If $\bs \nu$ is chosen so that $g(\bs \theta^{(l)}; \bs \nu)$ is improper, one can show that the posterior distribution is also improper \citep{maclachlanfinite, wasserman2000asymptotic}. For that reason, noninformative or weakly-informative base measures have been implemented with modifications of Jeffrey's prior \citep{diebolt1994estimation, rubio2014inference}, noninformative hierarchical priors \citep{grazian2018jeffreys}, and global reparameterizations \citep{robert1995reparameterization, kamary2018weakly}. Another strategy is to calibrate $\bs \nu$ via the data itself to create a data-based prior, see \cite{raftery1995hypothesis} and \cite{wasserman2000asymptotic} for applications of such methodology to the priors in a GMM.  

We will focus on the case when the data are modeled as a GMM with diagonal covariance. The cluster specific parameters are $\bs \theta^{(l)} = (\bs \mu^{(l)}, \Sigma^{(l)})$, where $\bs \mu^{(l)} = (\mu_1^{(l)}, \dots, \mu_p^{(l)})$ and $\Sigma^{(l)} = \tx{diag}(\sigma_1^{(l)2}, \dots, \sigma_p^{(l)2})$, or $\theta_j^{(l)} = (\mujl, \sigmajlsq)$. The data are generated by 
\begin{equation} \label{eq:yij-likelihood}
     (y_{ij} \mid c_i = l, \mujl, \sigmajlsq)  \sim \N(\mujl, \sigmajlsq), \tx{ for } j=1, \dots, p;
 \end{equation}
with $\mujl$ being the mean, or \textit{center}, and $\sigmajlsq$ being the variance of feature $j$ in cluster $l$, for $l=1,\dots,L$.  The natural choice for $\pi(\mujl, \sigmajlsq)$ is the Gaussian-inverse-Gamma base measure because it is conjugate to the normal likelihood, leading to a simple Gibbs sampler for $\pi(\bs c \mid \bs y)$. This choice of base measure has several pitfalls in practice, including the aforementioned ambiguity for the choice of hyperparameters and degenerate clustering behavior in high dimensional regimes \citep{chandra2023escaping}. In this article, we present an alternative prior for \eqref{eq:yij-likelihood} that is computationally tractable but favors interpretability in clustering. Similar to the approach of \cite{richardson1997bayesian}, we use expert prior information to construct the prior for $(\mujl, \sigmajlsq)$, but rather than setting a different weakly-informative prior for each cluster, we collate the information into a simple base measure. 

\subsection{Assumptions and Prior on the Cluster Centers}
Suppose that for variable $j$, $y_{ij} \in \chi_j \subset \R$ for all individuals $i=1,\dots,n$. The meaningful regions of $\chi_j$ are a set of subsets $\mathcal{M}_j = \lb \mathcal{M}_j^{(1)}, \dots, \mathcal{M}_{j}^{(K_j)} \rb$, so that $\bigcup_{k=1}^{K_j} \M_j^{(k)} = \chi_j$ and the event that $y_{ij} \in \M_j^{(k)}$ has a concrete meaning for the investigator. Typically, $\M_j$ is a set of disjoint intervals that comprise $\chi_j$. For instance, in the case where $\M_j$ reflects diminished, neutral, and elevated levels of variable $j$, we have that $K_j=3$, $\M_j^{(k)} = [a_j^{(k)}, b_j^{(k)}]$, and $a_j^{(1)}<b_j^{(1)}\leq a_j^{(2)} < b_j^{(2)} \leq a_j^{(3)} < b_j^{(3)}$. In medical applications, knowing that $y_{ij} \in \M_j^{(k_j)}$ for $j=1, \dots, p$ gives the clinician a clear interpretation in terms of individual $i$'s overall health, which can then be used to inform decisions such as treatment plans. We discuss alternative examples of MRs that arise in medicine in the Supplement.

We assume that the data are distributed as a GMM with diagonal covariance, such as in \eqref{eq:yij-likelihood}, with $0<L<\infty$ and $\bs \psi \sim \tx{Dir}(\gamma/L, \dots, \gamma/L)$, where $\bs \psi = (\psi_1, \dots, \psi_L)$ are the mixture weights, i.e. $\psi_l = \pi(c_i = l \mid \psi_l)$. $L$ is set to be a large fixed number, such as $L=10$ or $L=50$, so as to 
capitalize on the theoretic guarantees of such overfitted mixture models for determining the number of clusters \citep{rousseau2011asymptotic, van2015overfitting}. We set independent prior densities for $\mujl$ and $\sigmajlsq$, i.e. $\pi(\mujl, \sigmajlsq) = \pi(\mujl)\pi(\sigmajlsq)$. For the variance, we opt for the conjugate prior $1/\sigmajlsq \sim \tx{G}(\lambda_j, \beta_j)$, where $\tx{G}$ is the gamma distribution, with $\lambda_j$ and $\beta_j$ chosen so that $\pi(\sigmajlsq)$ is weakly-informative, though our approach is compatible with any variance prior, such as the hierarchical model given in \cite{richardson1997bayesian}. 

We now turn towards specifying a novel prior for $\mujl$ based on careful consideration of the MRs. For variable $j$ in cluster $l$, we assume that, a priori,
\begin{equation} \label{eq:CLAMR-prior-general}
    \mu_j^{(l)} \sim \sum_{k=1}^{K_j} \phi_j^{(k)} h_j^{(k)},
\end{equation}
where $0<\phi_j^{(k)}<1$, $\sum_{k=1}^{K_j}\phi_j^{(k)}=1$, and $h_j^{(k)}$ are continuous probability distributions with support on $\M_j^{(k)}$. The shape of the component distributions in \eqref{eq:CLAMR-prior-general} are determined using expert prior knowledge so that $h_j^{(k)}$ is itself an informative distribution on the $k$th MR. The resulting density function $\pi(\mujl)$ is multimodal, with modes contained inside each meaningful region. Our prior choice is a base measure for the centers, i.e. $\pi(\mujl) = g(\mujl; \bs \nu)$ with hyperparameters $\bs \nu = (\phi_j^{(1)}, \dots, \phi_j^{(K_j)}, h_j^{(1)}, \dots, h_j^{(K_j)})$. We refer to \eqref{eq:CLAMR-prior-general} as the CLAMR prior for cluster centers. Importantly, CLAMR is a prior for the cluster-specific parameters and not an assumption about the distribution of the data. This distinguishes the approach from mixtures of mixtures that model the within-cluster distributions as GMMs \citep{malsiner2017identifying}. 

The interpretation of \eqref{eq:CLAMR-prior-general} with respect to the data generating process can be better understood after the introduction of \textit{labels} $s_j^{(l)} \in \lb 1, \dots, K_j \rb$, with the property that the prior for $\mujl$ is equivalent to sampling
\begin{align} \label{eq:CLAMR-prior-labels}
    (\mu_j^{(l)} \mid s_j^{(l)} = k) & \sim h_j^{(k)}; & \pi(s_j^{(l)} = k) = \phi_j^{(k)};
\end{align}
such as in the typical mixture likelihood in \eqref{eq:yij-likelihood}. The height of each mode in \eqref{eq:CLAMR-prior-general} is controlled by $\bs \phi_j = (\phi_j^{(1)}, \dots, \phi_j^{(K_j)})$. If $\phi_j^{(k)} \approx 1$ for some $k$, then all cluster centers will concentrate inside the $k$th MR.
To model uncertainty in $\bs \phi_j$, we assume that $\bs \phi_j \sim \tx{Dir}(\rho_j/K_j, \dots, \rho_j/K_j)$ where $\rho_j > 0$. The smaller the value of $\rho_j$, the more likely it is that the cluster centers in feature $j$ are all representative of the same MR {\em a priori}. 

The generating process of the centers can be interpreted in terms of a \textit{profile matrix} $\s = (s_j^{(l)})_{l,j}$. Sampling \eqref{eq:CLAMR-prior-labels} associates the $l$th cluster center with a $p$-dimensional vector of MRs, $\bs s^{(l)} = (s_1^{(l)}, \dots, s_p^{(l)})$, which we refer to as the clinical \textit{profile} of cluster $l$. The profiles provide a concise medical description of the cluster centers. If, say, $p=3$ and the MRs correspond to expression levels, then $\s^{(l)} = (2,1,3)$ implies that cluster $l$ is characterized by neutral values of feature one, diminished measurements of feature two, and elevated levels of feature three. Since we expect label-switching to occur, $\s$ is identifiable up to permutations of the rows. The columns of $\s$, denoted $\s_j = (s_j^{(1)}, \dots, s_j^{(L)})$, are groupings of the $L$ clusters into the $K_j$ MRs, and give an idea on the effect of feature $j$ on heterogeneity between the clusters. We will expand on this specific property in the following section. 

\begin{figure}
    \centering
\begin{tikzpicture}[roundnode/.style={circle, draw=blue!60, fill=blue!5, thick, minimum size=3mm}, squarednode/.style={rectangle, draw=red!60, fill=red!5, thick, minimum size=5mm},box/.style = {draw,black,inner sep=10pt,rounded corners=5pt}, node distance = {15mm}]
        \node[squarednode] (y) {$y_{ij}$};
        \node[roundnode] (mu) [below left = of y] {$\mujl$};
        \node[roundnode] (sigma) [below right = of y] {$\sigmajlsq$};
        \node[roundnode] (c) [above = of y] {$c_i=l$};
        \node[roundnode] (xi) [below = of mu] {$\xijk$};
        \node[roundnode] (tau) [below right = of mu] {$\taujksq$}; 
        \node[roundnode] (lambda) [below = of sigma] {$\lambda_j$};
        \node[roundnode] (beta) [below right = of sigma] {$\beta_j$};
        \node[roundnode] (sj) [below left = of mu] {$s_j^{(l)}=k$};
        \node[roundnode] (gamma) [above = of c ] {$\gamma$};
        \node[roundnode] (rho) [below = of sj] {$\rho_j$};
        \draw[->] (c)--(y);
        \draw[->] (gamma)--(c);
        \draw[->] (mu)--(y);
        \draw[->] (sj)--(mu);
        \draw[->] (xi)--(mu);
        \draw[->] (tau)--(mu);
        \draw[->] (rho)--(sj);
        \draw[->] (sigma)--(y);
        \draw[->] (lambda)--(sigma);
        \draw[->] (beta)--(sigma);
        \node[box, fit=(c)(gamma)] {} ;
        \node[box, fit=(mu)(sj)(xi)(tau)(rho)] {} ;
        \node[box, fit=(sigma)(lambda)(beta)] {} ;
    \end{tikzpicture}
    \caption{The data generating process according to the CLAMR prior after marginalizing out the prior distributions for $\bs \psi$ and $\bs \phi_j$. The observation $y_i$ is allocated into cluster $l$ with probability determined by $\gamma$, with mean and variance equal to $\mujl$ and $\sigmajlsq$ for variable $j$. The cluster center is defined by the cluster's profile $\s^{(l)}$, with $\xijk$ and $\taujksq$ chosen so that the kernels have support on the meaningful regions.
    The cluster-specific variance is drawn from an inverse-gamma conjugate prior.}
    \label{fig:dag}
\end{figure} 

In practice, the component kernels $h_j^{(k)}$ can be any continuous probability distribution, but a natural choice is the location-scale Gaussian mixture across the MRs,
\begin{equation} \label{eq:CLAMR-prior-GMM}
    \mujl \sim \sum_{k=1}^{K_j} \phi_j^{(k)} \N(\xi_j^{(k)}, \tau_j^{(k)2}),
\end{equation}
where $\xijk \in \M_j$ and $\tau_j^{(k)2} > 0$. The data generating process of $\y$ using Gaussian components in the CLAMR prior is visualized with a directed acyclic graph (DAG) in Figure \ref{fig:dag}. The three main pieces of the generation of $y_{ij}$ are represented by three boxes, with each box being independent of the other a priori.  
The box on the bottom left of the DAG details the prior on the cluster center. 

The component means and variances of \eqref{eq:CLAMR-prior-GMM} must be chosen carefully so that the bulk of the probability mass of each Gaussian kernel are non-overlapping. Clearly, $\xijk$ must be some real number contained inside $\M_j$. In the case where $\M_j^{(k)} = [a_j^{(k)}, b_j^{(k)}]$, $\xi_j^{(k)}$ is any point in that interval. A natural default version of \eqref{eq:CLAMR-prior-GMM} is the following specification based on controlling the symmetric tails of the Gaussian kernel,
\begin{align} \label{eq:CLAMR-prior-default}
    \xijk & = \frac{a_j^{(k)} + b_j^{(k)}}{2} & \taujksq = \lb  \frac{b_j^{(k)} - a_j^{(k)}}{2\Phi^{-1}\left( \frac{1+\omega}{2} \right)} \rb^2,
\end{align}
where $0 < \omega < 1$ is a constant.  
Note that $\xijk$ is the midpoint of $\M_j^{(k)}$, similar to the prior specification in \cite{richardson1997bayesian}. The free parameter $\omega$ controls the amount of probability mass that overlaps between MR $k$ and MRs $k-1$ and $k+1$. One can see that $\pi(\mujl \in \M_j^{(k)} \mid s_j^{(l)} = k) = \omega$. Increasing $\omega$ to $1$ will increase the concentration of each kernel towards their midpoint, causing $\pi(\mujl) \to \sum_{k=1}^{K_j} \phi_j^{(k)} \delta_{\xijk}$. Hence, a sensible choice is to set $\omega = 0.95$ so that $95\%$ of the probability density in each kernel is contained within their respective region and centered at the midpoint. If investigators do not wish to place $\xijk$ as the midpoint of $\M_j^{(k)}$, they can choose $\taujksq$ using a similar criteria to \eqref{eq:CLAMR-prior-default} to ensure that $\M_j^{(k)}$ contains $95\%$ of the probability density of component $k$. The framework can also be extended to multimodality within MRs by setting $h_j^{(k)}$ to be mixture of Gaussian kernels, combining CLAMR with mixtures of mixtures \citep{malsiner2017identifying}.

\begin{figure}[t]
    \centering
    \includegraphics[scale=.55]{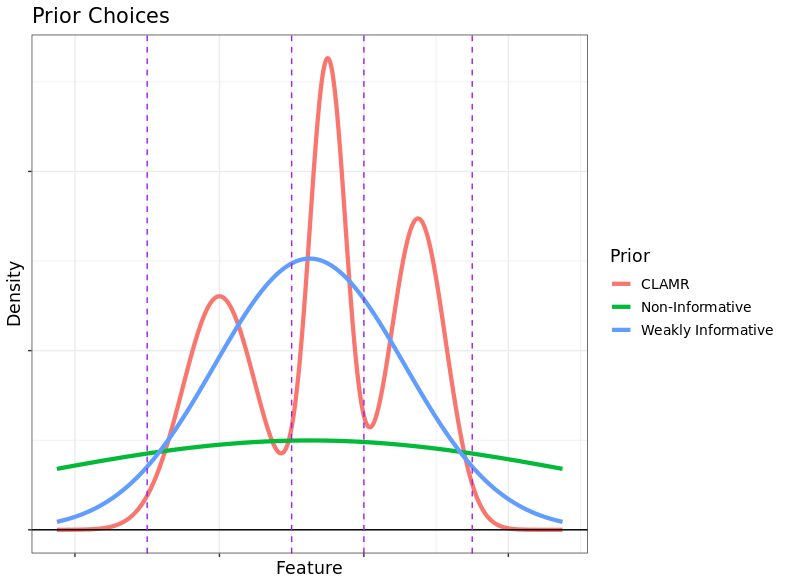}
    \caption{Possible prior choices for the cluster center when the MRs are given by $\M = \lb [-3,-1], [-1,0],[0,1.5] \rb$ and $\chi_j = [-3,1.5]$. The non-informative prior is flat on $\chi_j$, whereas the weakly-informative choice is concentrated at the midpoint of $\chi_j$. The CLAMR prior has hyperparameters given by \eqref{eq:CLAMR-prior-default}, where each component of the mixture is a non-overlapping Gaussian with mean at $(a_j^{(k)} + b_j^{(k)})/2$. In this specific example, the prior probability under CLAMR of observing a cluster center in each MR is $1/3$, though that need not be the case in general.}
    \label{fig:prior-choices}
\end{figure}

An example comparison between the CLAMR prior, the weakly-informative case (as mentioned in \cite{richardson1997bayesian}), and the noninformative case is given in Figure \ref{fig:prior-choices} when $\chi_j = [-3,1.5]$, $\M_j = \lb [-3,-1], [-1,0], [0,1.5] \rb$, and the MRs are expression levels. The CLAMR prior favors clusters with centers located at the midpoint of each MR, and the prior probability of the MRs is controlled by $\bs \phi_j$. For this visualization, $\bs \phi_j = (1/3,1/3,1/3)$, but the height of each mode can be lowered or raised by altering these values. In the case where $\phi_j^{(k)} \to 1$ for some $k$, the CLAMR prior in Figure \ref{fig:prior-choices} reduces to an informative prior on the $k$th MR. The weakly-informative prior, while leading to a proper posterior, favors neutral levels of feature $j$, and the prior probability of each MR is fixed. 

\subsection{Testing Feature Influence and Pre-Training}

The clinical profiles $\s$ will be useful for assessing the \textit{influence} of feature $j$ on $\c$. In phenotype analyses, investigators are interested in what features help explain the phenotypes and which variables do not. The intuition here is that some features tend to contribute more than others to meaningful differences between the groups. Suppose then, for example, that for some $k$, $s_j^{(l)} = k$ for all clusters $l=1, \dots, L$. This would mean that the profiles of each cluster are all identical for the $j$th variable. If $k=2$ and the MRs are expression levels, this would amount to concluding that individuals in all clusters tend to have neutral values of variable $j$. Therefore, variable $j$ on its own does not explain the heterogeneity between groups. 

We can formulate this intuition in terms of a Bayesian hypothesis test. Recall that for each cluster $l$ and variable $j$, there is a label $s_j^{(l)} \in \lb 1, \dots, K_j \rb$. In practice, there may be some clusters that are empty, so for the remainder of this section, we only focus on the clusters $C_l$ that are non-empty, i.e. $|C_l| \geq 1$. Without loss of generality and for ease of notation, we will assume that clusters $l=1, \dots, L_n \leq L$ are non-empty. The labels induce $p$ partitions of $[L_n] = \lb 1, \dots, L_n \rb$, denoted $ S_j = \lb S_j^{(1)}, \dots, S_j^{(K_j)} \rb$, where $S_j^{(k)} = \lb l : s_j^{(l)} = k, |C_l| \geq 1 \rb$, for $j=1, \dots, p$. That is, each $S_j^{(k)}$ is the set of cluster indices that are associated with the $k$th MR. For example, suppose there are $L_n=3$ clusters and $K_j=3$ MRs. Then by sampling \eqref{eq:CLAMR-prior-labels} there are three possibilities for the allocation of clusters to regions: all clusters are in the same MR ($S_j = \{[3]\}$), two clusters are in the same MR while one cluster is in a different MR ($S_j = \lb \{1,2\}, \{3 \} \rb$, $S_j = \lb \{1\}, \{2,3\} \rb$, or $S_j = \lb \{1,3\}, \{2 \} \rb$), or all clusters are in different MRs ($S_j = \lb \{1\}, \{2 \}, \{3 \} \rb$).

From our argument above, if $S_j = \{ [L_n] \}$, then the influence of variable $j$ on $\c$ is minimal because each cluster is located in the same MR. 
This motivates setting the null hypothesis to be $H_{0j}:S_j=\{ [L_n] \}$. 
However, in this article, we set the interval null for feature influence to be $H_{0j}:d(S_j, \{ [L_n] \}) < \epsilon_j$, where $\epsilon_j > 0$ is a positive constant and $d$ is a distance metric over the partition space. We implement our approach with $d$ set to be Binder's loss \citep{binder1978bayesian}, but other partition metrics exist that could be used instead, e.g. the Variation of Information (VI) \citep{meilua2007comparing}.

In accordance with formal Bayesian hypothesis testing, we focus on the prior and posterior probability of $H_{0j}:d(S_j, \{ [L_n] \}) < \epsilon_j$ versus $H_{1j}: d(S_j, \{ [L_n] \}) \geq \epsilon_j$. $H_{0j}$ is tested via computation of the Bayes factor ($\tx{BF}_j$), or the ratio of the marginal likelihoods of $\y$ given $H_{1j}$ and $H_{0j}$ \citep{jeffreys1935some, jeffreys1961theory, kass1995bayes}. When $\pi(H_{0j})=\pi(H_{1j})=1/2$, the Bayes factor can be further simplified to $\tx{BF}_j = \pi(H_{1j} \mid \y)/\pi(H_{0j} \mid \y) =  \lb 1 - \pi(d(S_j, \{ [L_n] \}) < \epsilon_j \mid \y) \rb / \pi(d(S_j, \{[L_n] \}) < \epsilon_j \mid \y)$, 
which is the ratio of the posterior probabilities of $H_{1j}$ and $H_{0j}$. 
We can conclude in favor of either $H_{0j}$ or $H_{1j}$ using the magnitude of $\tx{BF}_j$, e.g. the guidelines given in \cite{kass1995bayes}. The smaller $\tx{BF}_j$ is, the smaller the influence of feature $j$ on $\c$.
Note that $\pi(H_{0j})=1/2$ only occurs for specific values of $\rho_j$ and $\gamma$, so we fix $\gamma=1$, which favors a small number of clusters, then choose $\rho_j$ using Monte Carlo simulation.

In practice, testing feature influence can be implemented as a pre-training step in order to reduce dimension in $\bs y$ and identify relevant features for clustering. This involves first fitting the model, computing the Bayes factors, and removing the non-influential features. We then run the CLAMR model on the reduced set of variables and infer the clustering $\bs c$. In Section \ref{sect:indite} we implement our hypothesis testing scheme as a pre-training procedure to remove noisy features in the INDITe data.

\subsection{Computational Details}

We derive a simple Gibbs sampler for CLAMR, which is detailed in the Supplement. Once the $T$ posterior samples $\{ \c^{(t)} \}$ and $\{ \s^{(t)} \}$ are obtained, we summarize our findings using MCMC estimation. A key advantage of the Bayesian framework is the ability to characterize uncertainty in point estimation using the posterior distribution. In Bayesian cluster analysis, the point estimate $\c^* = (c_1^*, \dots, c_n^*)$ of $\c$ is a fixed clustering of $\y$ based on $\pi(\c \mid \y)$, and is derived by minimizing the posterior expectation of an appropriate clustering loss function. We minimize the Variation of Information (VI) loss \citep{meilua2007comparing}, which compares the entropy between $\c^*$ and $\c$ as well as their mutual information \citep{wade2018bayesian}. However, CLAMR can be implemented with other losses if desired, including Binder's pair-counting loss \citep{binder1978bayesian} and the robustifying loss FOLD \citep{dombowsky2024bayesian}. For uncertainty quantification, we compute a
credible ball or PSM using the MCMC samples. To test the influence of the $j$th feature, $\tx{BF}_j$ can be consistently estimated as $T \to \infty$ by $\pi(d(S_j, [L_n]) < \epsilon_j \mid \y) \approx (1/T)\sum_{t=1}^{T}\textbf{1}(d(S_j^{(t)}, [L_n^{(t)}]) < \epsilon_j)$, and is computed using post-processing on the samples $\{ \c^{(t)} \}$ and $ \{ \s^{(t)} \}$. 

\subsection{Cluster Interpretations}
The non-identifiability of cluster labels in Bayesian mixture models makes estimating the profiles $\bs s$ challenging without applying a relabeling algorithm. However, once a point estimate $\bs c^*$ is obtained, we can utilize these fixed labels to infer MR specifications. Denote $C^* = (C_1^*, \dots, C_M^*)$, where $C_m^* = \lb i: c_{i}^*=m \rb$ is the set of participants allocated to cluster $m$ by the point estimate $\bs c^*$. For participant $i$, define the quantity $\hat{s}_j^{(i)} = s_j^{(l)}$ if and only if $c_i=l$. Recall that $c_i$ is the mixture label of participant $i$ (which is random), whereas $c_i^*$ is the cluster label in the point estimate (which is fixed). We compute
\begin{equation} \label{eq:Delta}
   \Delta_j^{(m,k)} = \frac{1}{|C_m^*|}\sum_{i \in C_m^*} \textbf{1}_{\hat{s}_j^{(i)} = k} \tx{ for all } m \in [M], k \in [K_j], j \in [p].
\end{equation}
This quantity can be interpreted as the empirical probability of association with the $k$th MR in variable $j$ for the individuals allocated to cluster $m$ of $\bs c^*$.

We compute \eqref{eq:Delta} for every iteration in the Gibbs sampler, then summarize our findings using
\begin{equation} \label{eq:MR-specification-estimate}
    \Delta_j^{m,*} = \max_{k=1, \dots, K_j} \bar{\Delta}_j^{(m,k)} \tx{ and } s_j^{m,*} = \underset{k=1, \dots, K_j}{\tx{arg max}} \bar{\Delta}_j^{(m,k)}
\end{equation}
for all clusters $m$ in $\bs c^*$ and features $j$; here, $\bar{\Delta}_j^{m,j}$ is the Monte Carlo average of $\Delta_j^{(m,j)}$. We interpret $s_j^{m,*}$ as the MR most associated with the participants in the $m$th group of $\bs c^*$, with $\Delta_j^{m,*}$ as a measure of variation about this MR. Alternatively, uncertainty in MR assignment can be quantified by inspecting the posterior distributions of $\Delta_j^{(m,k)}$.

\section{Simulation Studies} \label{sect:simulations}

\subsection{Clustering Under Misspecification} \label{sect:sims-misspec}
We now evaluate CLAMR on synthetic data. We focus on the case in which the statistician has obtained expert knowledge on the MRs but the model is \textit{misspecified}. Our simulation scenario explores kernel misspecification, which in our case refers to the regime in which the true data generating process is a mixture model but with non-Gaussian components. Kernel misspecification is a key issue in model-based clustering because statistical methods tend to overestimate the number of clusters when there is a mismatch between the assumed kernels and the true ones. A number of methods have recently been proposed to robustify Bayesian clustering results to model misspecification, such as coarsening \citep{miller2018robust}, decision theoretic schemes \citep{dombowsky2024bayesian, buch2024bayesian}, and hierarchical component merging algorithms \citep{aragam2020identifiability, do2024dendrogram}. However, the effect of the prior distribution on robustifying inference has received little discussion in this nascent literature.

\begin{table}[ht]
\centering
\begin{tabular}{rrrrrrr}
  \toprule
 & $n$ & CLAMR & BGMM & EM & k-means & HCA \\ 
  \midrule
ARI & $100$ & 0.96 (0.10) & 0.99 (0.02) & 0.99 (0.03) & 0.32 (0.17) & 0.22 (0.17) \\ 
  & $500$ & 0.98 (0.05) & 0.98 (0.01) & 0.94 (0.06) & 0.35 (0.17) & 0.20 (0.15) \\ 
  & $750$ & 0.98 (0.04) & 0.97 (0.02) & 0.94 (0.05) & 0.34 (0.18) & 0.17. (0.15) \\ 
  & $1000$ & 0.98 (0.01) & 0.97 (0.02) & 0.93 (0.06) & 0.34 (0.16) & 0.17 (0.12) \\ 
   \midrule
 $\hat{L}$ & $100$ & 3.14 (0.47) & 3.34 (0.55) & 3.14 (0.38) & 3.00 (0) & 3.00 (0) \\ 
  & $500$ & 4.41 (1.00) & 4.60 (0.89) & 3.90 (0.63) & 3.00 (0) & 3.00 (0) \\ 
  & $750$ & 4.64 (1.08) & 4.93 (0.89) & 4.10 (0.58) & 3.00 (0) & 3.00 (0) \\ 
  & $1000$ & 5.07 (1.01) & 5.44 (0.92) & 4.27 (0.49) & 3.00 (0) & 3.00 (0) \\  
\end{tabular}
\caption{Average adjusted Rand index (ARI) with $\tilde{\bs c}$ and number of clusters ($\hat L$) for CLAMR, the BGMM, EM, k-means, and HCA applied to $100$ independent replications for each $n \in \lb 100, 500, 750, 1000 \rb$. Both k-means and HCA have the number of clusters fixed at $L=3$, the true value. Standard deviations of the metrics across the replications are given in brackets.}
\label{table:sims-results}
\end{table}

We find that CLAMR can reduce the over-clustering tendency of misspecifed models, leading to more accurate estimation and a fewer number of clusters. The synthetic data are comprised of $p=6$ features with $K_j = 3$ true MRs each, which are assumed to be known. The number of true clusters is set to $\tilde L = 3$ and we simulate the true within-cluster center parameters, denoted $\tilde{\mu}_j^{(l)}$, from the CLAMR prior \eqref{eq:CLAMR-prior-GMM}, where the true profile matrix $\tilde{\bs s}$ is given by
\begin{equation*}
    \tilde{\bs s} = 
    \begin{pmatrix}
        1 & 3 & 1 & 2 & 1 & 3 \\
        1 & 3 & 2 & 1 & 1 & 3 \\
        1 & 3 & 3 & 3 & 1 & 3\\
    \end{pmatrix}.
\end{equation*}
Observe that this implies that only features $3$ and $4$ will exhibit meaningful between-cluster variation. To replicate our application to the INDITe data, we make each feature exhibit a different scale. For example, the MRs of feature $1$ are $\lb [-1,1], [1,2], [2,4] \rb$, and the MRs for feature $6$ are $\lb [0,10], [10,30], [30,200] \rb$. True cluster labels $\tilde c_i$ are simulated uniformly across the integers $[\tilde L].$ The cluster-specific scales $ \tilde \sigma_j^{(l)}$ are chosen so that, for the $l$th true cluster, the simulated data are contained within the MR given by the corresponding entry in $\tilde{\bs s}$. This calibration uses the true data generating process, given by $y_i \mid \tilde c_i = l \sim t(\tilde \nu,  \tilde \mu_j^{(l)}, \tilde \sigma_j^{(l)}) $, where $t(\nu, \mu, \sigma)$ denotes the Student's t-distribution with $\nu$ degrees of freedom, center $\mu$, and scale $\sigma$. For our simulations, we fix $\tilde \nu = 5$, meaning that the true kernels have significantly heavier tails than the Gaussian distribution. We vary $n \in \lb 100, 500, 750, 1000 \rb$ and simulate $100$ independent replications for each sample size. For a given simulated dataset $\bs y$, we fit a GMM with the CLAMR prior and default hyperparameters \eqref{eq:CLAMR-prior-default} with $\omega = 0.95$ and $\gamma=1$. We compare the results to clusterings yielded by a standard Bayesian GMM with weakly-informative priors (BGMM), an implementation of the EM algorithm, complete linkage hierarchical clustering (HCA), and k-means. Details on hyperparameter choice for all approaches are given in the Supplement. For both CLAMR and the BGMM, a clustering point estimate is selected by minimizing the VI loss. 

Results for the simulation study are displayed in Table \ref{table:sims-results}, which shows the average adjusted Rand index (ARI) \citep{rand1971objective,hubert1985comparing} between each point estimate and the true cluster labels $\tilde{\bs c}$, as well as the number of clusters in the point estimates $(\hat L)$. For $n>100$, CLAMR achieves the highest ARI across all approaches. In addition, CLAMR consistently reduces the number of clusters on average in comparison to the standard BGMM. The EM algorithm excels at inferring the number of clusters, but is notably less accurate at inferring the true cluster membership, as evidenced by the smaller values of ARI. Despite being fixed at the true number of clusters, k-means and HCA completely fail at recreating the true cluster structure. The supplemental material details a similar simulation study in which the model is well-specified, i.e., we simulate the data from Gaussian distributions.

\subsection{Impact of MR Choice} \label{sect:sims-no-MRs}

\begin{table}[ht]
\centering
\begin{tabular}{rrrrrrr}
  \toprule
 & $n$ & CLAMR & BGMM & EM & k-means & HCA \\ 
  \midrule
ARI & $100$ & 0.98 (0.08) & 0.98 (0.08) & 0.99 (0.03) & 0.71 (0.20) & 0.67 (0.21) \\
  & $500$ & 0.99 (0.01) & 0.99 (0.01) & 0.96 (0.05) & 0.77 (0.19) & 0.62 (0.24) \\  
  & $750$ & 0.99 (0.01) & 0.98 (0.01) & 0.92 (0.07) & 0.75 (0.21) & 0.61 (0.24)\\ 
  & $1000$ & 0.98 (0.04) & 0.98 (0.02) & 0.91 (0.06) & 0.79 (0.18) & 0.62 (0.22) \\
   \midrule
 $\hat{L}$ & $100$ & 3.09 (0.40) & 3.06 (0.34) & 3.15 (0.44) & 3.00 (0)  & 3.00 (0) \\ 
  & $500$ & 4.09 (0.84) & 4.20 (0.79) & 3.77 (0.69) & 3.00 (0) & 3.00 (0) \\ 
  & $750$ & 4.43 (0.99) & 4.49 (0.81) & 4.24 (0.74) & 3.00 (0) & 3.00 (0) \\
  & $1000$ & 4.92 (0.93) & 5.00 (0.89) & 4.47 (0.61) & 3.00 (0) & 3.00 (0) \\ 
\end{tabular}
\caption{ARI with $\tilde{\bs c}$ and $\hat L$ for CLAMR, the BGMM, EM, k-means, and HCA applied to $100$ independent replications for each $n \in \lb 100, 500, 750, 1000 \rb$ (brackets indicate standard deviation).}
\label{table:sims2-results}
\end{table}

In this simulation study, we evaluate the impact of specifying MRs when analyzing a dataset with no true MR structure. That is, we do not assume that the true cluster centers are simulated from a GMM with approximately non-overlapping components. Instead, we generate synthetic data $\bs y$ from a standard mixture model, i.e. $\bs y_i \sim \sum_{\tilde l = 1}^{\tilde L} t(\tilde \nu, \tilde \mu_j^{(l)}, \tilde \sigma_j^{(l)2})$. We simulate the true cluster variances in the same manner that we sampled the true cluster scales in the previous simulation study. However, the cluster centers are sampled in a manner without MR structure, i.e. $\tilde \mu_j^{(l)} \sim \N(\xi_j, \tau_j^2)$, where $\xi_j$ and $\tau_j^2$ are chosen so that 95\% of the Gaussian density encompasses \textit{ all} of the specified MRs. Continuing with our earlier example, if the MRs of feature 1 are $\lb [-1,1], [1,2], [2,4] \rb$, then $(\xi_j, \tau_j^2)$ are chosen according to \eqref{eq:CLAMR-prior-default} where $K_j=1$, $a_j^{(1)} = -1$, and $b_j^{(1)} = 4$. In the context of our application, this is a setting in which approximate bounds for the features are known. As before, we set the degrees of freedom to be $\tilde \nu = 5$. All other hyperparameters and settings for CLAMR, the BGMM, EM, k-means, and HCA are kept the same as in the previous simulation study.

We display the results in Table \ref{table:sims2-results}. Interestingly, we find that CLAMR can improve over BGMM in both accuracy and estimating the number of clusters, despite the MRs having no concrete bearing on the data generating process aside from approximate bounds on the centers. As we observed previously, both CLAMR and the BGMM have higher ARI than other methods, though the EM algorithm is generally more robust in this case to over-clustering the data. We do see improvement in the performance of k-means and HCA in this setting, but their average ARI scores are markedly lower than those of model-based methods. 

\section{Sepsis Phenotypes in Northern Tanzania} \label{sect:indite}

\subsection{Data and Pretraining}

Our main interest is using CLAMR to derive interpretable phenotypes from a cohort of sepsis patients. While the original dataset consists of a vast collection of variables, we opt for features that have been used in SENECA, which we enumerate in Table \ref{table:features-and-bfs}. The clinical variables can be categorized as belonging to four distinct classes: clinical signs, markers of inflammation in the body, markers of organ dysfunction, and additional immunological markers. The original INDITe cohort consisted of febrile patients that are not necessarily sepsis positive, so we only include individuals with systemic inflammatory response syndrome (SIRS) score greater than or equal to $2$ in order to derive phenotypes that correspond to sepsis definition, ultimately resulting in $n=265$. The following features have missing observations: bilirubin (0.023\%), platelets (0.011\%), BUN (0.011\%), WBC (0.011\%), and bicarbonate (0.008\%). Based on our knowledge of the data collection process, we assume that these variables are missing completely at random (MCAR), and we impute these entries using the Gibbs sampler. We also have information on the in-patient outcome (e.g., death), which we do not use for clustering but will refer to in our interpretations. 

 MRs correspond to expression levels specified by our clinical science collaborators, who have expert knowledge of the patient population context in northern Tanzania. Thirteen of the characteristics have expression levels written in terms of D, N, and E ($K_j=3$), while the remaining two are expressed in D and N or N and E ($K_j=2$). For example, systolic blood pressure, which is measured in mm Hg, has the following MRs: $\tx{D}=[65,111]$, $\tx{N}=[111,220]$, and $\tx{E}=[220,300]$. In contrast, creatinine, measured in ml / dL, has only two MRs: $\tx{N}=[0,90]$ and $\tx{E}=[90,250]$. These cutoffs are unique to northern Tanzania and are non-overlapping. For more general implementations of the CLAMR algorithm, overlapping MRs may be considered, which we discuss in Section \ref{sect:posterior-predictive} and the Supplement. However, these D/N/E MRs are the most straightforward MR specifications for our clinical collaborators to elicit, and their clinical relevance makes them sensible choices for creating subgroups that could be used for treatment design. These MRs are then used to define the hyperparameters $(\xijk, \sigmajlsq)$ through \eqref{eq:CLAMR-prior-default} with the overlap parameter set to $\omega=0.95$. No other preprocessing is applied to the data, and features are kept in their original units for the analysis. 

Before computing the clustering point estimate, we implement a pretraining phase to select influential features for clustering. We set $\gamma=1$ and choose $\rho_j  = (1.1)\textbf{1}_{K_j=2}+(0.7)\textbf{1}_{K_j=3}$, which corresponds to $\pi(d(S_j, [L_n])<0.1) = 1/2$. The resulting Bayes factors for all variables are displayed in the third column of Table \ref{table:features-and-bfs}, with bolded script indicating features that have substantial evidence of clustering influence \citep{kass1995bayes}. We select $p=4$ of the original 15 features for our analysis, corresponding to AST, bilirubin, BUN, and creatinine, all of which are organ dysfunction markers. We then calculate the point estimate using the reduced set of features.

\begin{table}[]
\centering
\begin{tabular}{lll}
\toprule
  Feature  & Type & $\tx{BF}_j$    \\
\midrule
Temperature         & \multirow{4}{7em}{Clinical Signs} & 6.392    \\
Respiratory Rate       &  & 0.024    \\
Pulse Rate         &    & 0.402    \\
Systolic Blood Pressure (SBP)  &    & 2.765    \\
\midrule
White Blood Cell Count (WBC)    & \multirow{2}{7em}{Inflammation Markers}   & 14.553  \\
C-Reactive Protein (CRP)  &  & 2.636  \\
\midrule
Aspartate Transaminase (AST)         &  \multirow{5}{7em}{Organ Dysfunction Markers}   & \textbf{250.736 }   \\
Bilirubin         &    & \textbf{169.488}   \\
Blood Urea Nitrogen (BUN)  &    & \textbf{12607.696  }  \\
Creatinine &    & \textbf{45.208}   \\
Platelets   &  & 3.500  \\
\midrule
Glucose   &  \multirow{4}{7em}{Additional Markers} & 0.661    \\
Sodium     &  & 1.746 \\
Bicarbonate      &    & 6.139  \\
Albumin     &  & 4.600  \\
\bottomrule
\end{tabular}
\caption{The complete 15 variables from the INDITe data with accompanying Bayes factors from the pre-training phase. Features with Bayes factors indicating strong evidence against $H_0$ are indicated by bolded script.}
\label{table:features-and-bfs}
\end{table}

\subsection{Clustering Point Estimate and Interpretations}

\begin{table}
\centering
\setlength{\tabcolsep}{5pt}
\begin{tabular}{llllllll}
\toprule
   & Overall & $C_1^*$   & $C_2^*$   & $C_3^*$   & $C_4^*$   & $C_5^*$ & $C_6^*$    \\
\midrule
Size    & 264 & 87   & 49    & 44    & 39    & 27   & 18     \\
$\Pr$(Female) & 0.485  & 0.690 & 0.388 & 0.409 & 0.436 & 0.185 & 0.500    \\
$\Pr$(HIV+)  & 0.383   & 0.414 & 0.306 & 0.523 & 0.359 & 0.185 & 0.444  \\
$\Pr$(Adv. HIV) & 0.258  & 0.241 & 0.224 & 0.386 & 0.231 & 0.148 & 0.333 \\
$\Pr$(Malaria) & 0.098 & 0.092 & 0.163 & 0.045 & 0.051 & 0.185 & 0.056 \\
$\Pr$(Deceased) & 0.125 & 0.057 & 0.122 & 0.182 & 0.179 & 0.148 & 0.167   \\
Avg. Age (SD) & 42.8 (17) & 37 (14.2) & 47.5 (17.5) & 38.2 (12.1) & 48.5 (19) & 40.4 (16.8) & 60.2 (16.6) \\
\midrule
BUN & $-$ & N (0.98) & N (0.81) &   N (0.95)  &  E (0.68)   & E  (0.89) &  E (0.98)  \\
Creatinine & $-$ & N (0.98) &  N (0.91) &  N (0.95) &   E (0.54) &  N (0.63) &  E (0.91)\\
AST & $-$ & N (0.93)  &  N (0.93)  & E (0.90) &  N  (0.84) & E (0.93) &   N  (0.87)   \\
Bilirubin & $-$ & D  (0.89) & N  (0.86) &  D (0.89)  &  D (0.66)  &  N (0.89) &   N (0.77)\\

\midrule
\end{tabular}
\caption{Basic demographic information consisting of sex, HIV infection, advanced HIV status, malaria status, inpatient outcome, and age measured within each cluster, with the overall numbers included for comparison. In addition, estimated profiles $s_j^{m*}$, with $\Delta_j^{m,*}$ in parentheses, for each cluster and feature are included.}
\label{table:demographic-information}
\end{table}

\begin{figure}[ht]
    \centering
    \includegraphics[scale=0.52]{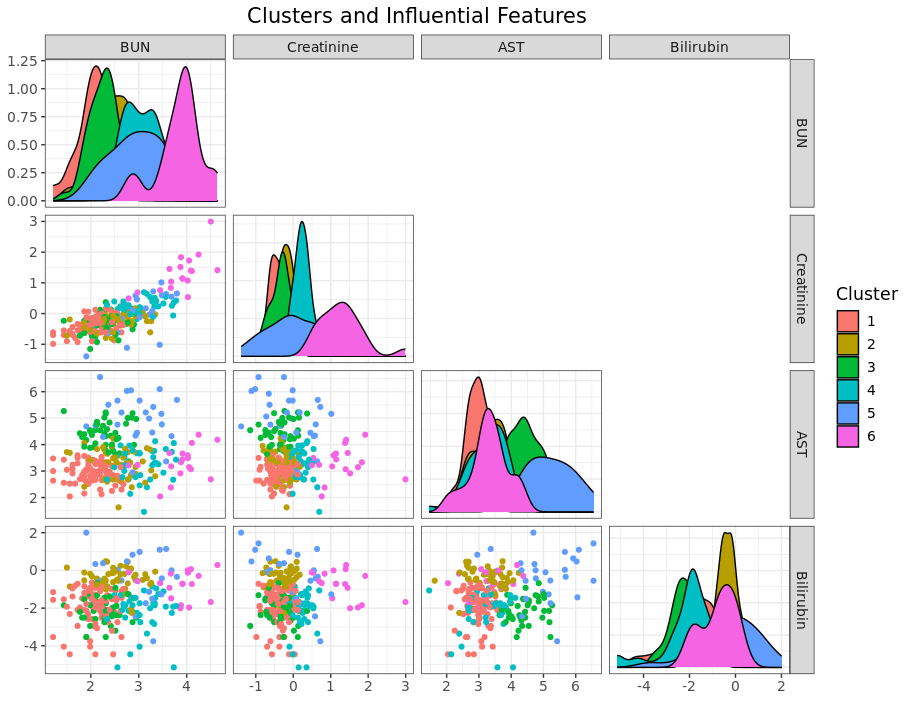}
    \caption{Pairs plots for the influential features BUN, creatinine, AST, and bilirubin, where colors indicate clusters. All features are log-transformed.}
    \label{fig:influential-pairs}
\end{figure}

\begin{figure}
    \centering
    \includegraphics[scale=0.6]{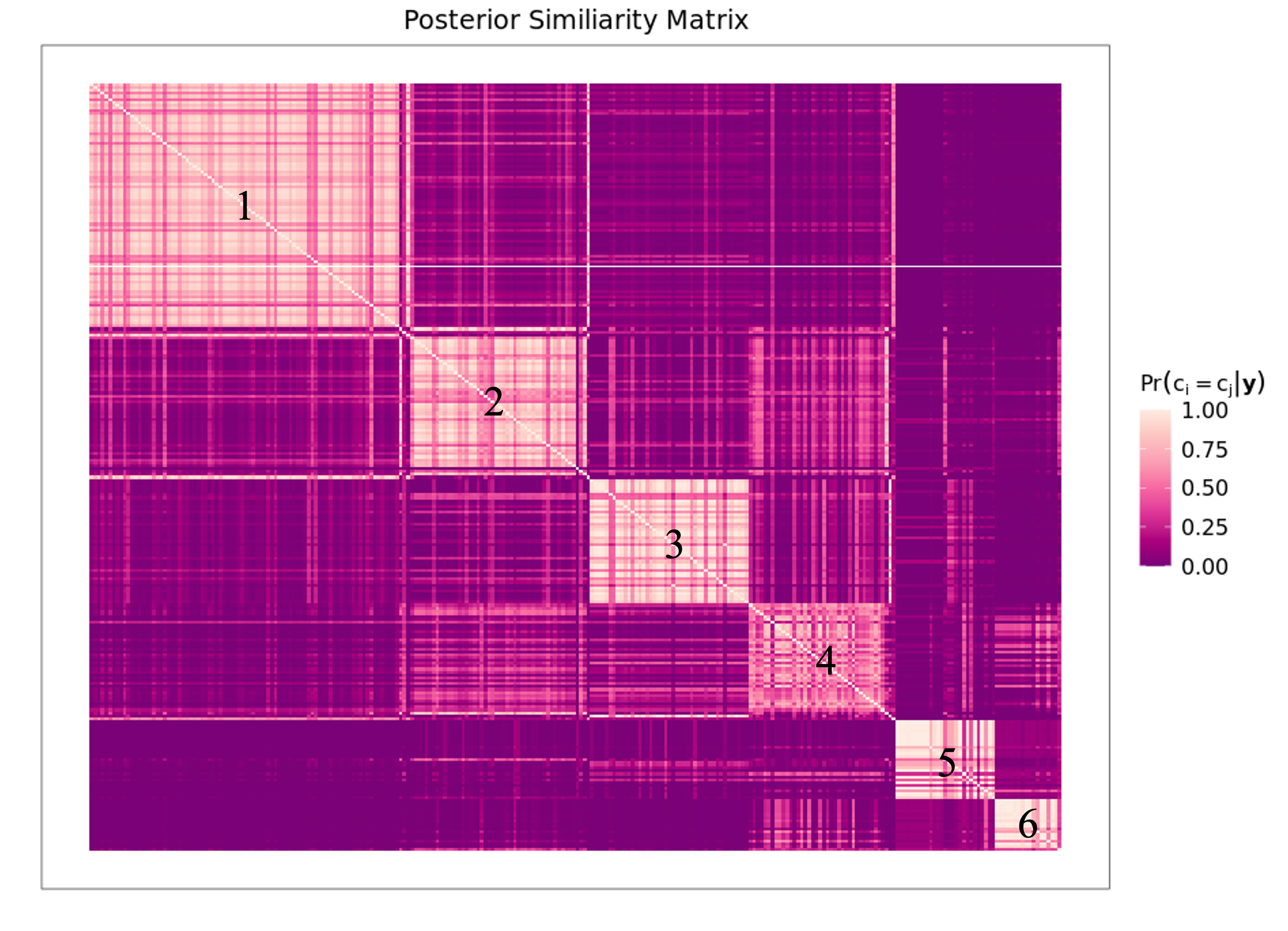}
    \caption{Posterior similarity matrix (PSM) for the INDITe data, ordered by point estimate membership.}
    \label{fig:psm}
\end{figure}

We compute the point estimate $\bs c^*$ by minimizing the VI loss using the MCMC samples from $\pi(\bs c \mid \bs y)$. We also calculate the PSM for the INDITe data to express uncertainty in our point estimate. The point estimate consists of $7$ clusters, however, the seventh cluster contains only one individual who has abnormally high levels of bilirubin, so we focus on the $6$ clusters derived from the remaining $264$ participants. We display a general summary of relevant clinical characteristics among participants, including the proportion of females, persons living with HIV, malaria infection, in-hospital mortality, and age in Table \ref{table:demographic-information} stratified by cluster; we provide a pairs plot of the influential features in Figure \ref{fig:influential-pairs}. For convenience, the clusters are labeled by decreasing size. No cluster contains the majority of participants, and the smallest cluster is cluster 6 which comprises just 18 individuals (or 7\%) from the overall cohort. In Figure \ref{fig:influential-pairs}, observe that cluster $6$ displays multimodality for some features (e.g., the bilirubin measurements). This is largely due to the fact that the sample size for this cluster is small relative to the size of the cohort, as it includes fewer than 7\% of the participants.

To aid interpretation, we compute the estimated MR specifications in \eqref{eq:MR-specification-estimate}, while also inspecting descriptive statistics for the clusters post hoc. Participants in cluster 1 are generally female (69\%), have the best inpatient outcome (5.7\% mortality), and have the lowest average age in comparison to the other clusters (37). These participants tend to have neutral feature expression with the exception of bilirubin,  which tends to exhibit diminished values. In comparison, cluster 2 is majority male (61\%), has a lower proportion of participants living with HIV (31\% in comparison to 41\% for cluster 1), and has more than double the mortality (12\%). Cluster 2 is also generally associated with neutral feature expression, though it is more inclined towards higher levels of BUN, bilirubin, and AST than cluster 1. Cluster 3 is the only group for which the majority of patients live with HIV (52\%, the overall incidence is 38\%), and 74\% of these patients have a percentage of CD4 indicative of advanced HIV. This group has the lowest proportion of subjects with malaria (4. 5\%, the prevalence among all $264$ is 9. 8\%), the participants are generally male, younger (average age 38.2 years) and have elevated levels of AST. Cluster 4 is characterized by elevated BUN and creatinine levels(though the $\Delta_j^{4,*}$ values of these features are both $<0.7$), but neutral AST levels and generally diminished bilirubin. Clusters 3 and 4 are similar in size (44 and 39 participants, respectively), and have the joint highest mortality rate (18.2\% and 17.9\%, respectively). The highest levels of AST are observed in cluster 5, which is also associated with neutral bilirubin and the highest proportion of men (81. 5\%). Furthermore, this group has the lowest proportion of participants living with HIV (18. 5\%), although 80\% of these individuals have advanced HIV. Cluster 6, the smallest cluster, is characterized by markedly high levels of both BUN and creatinine, but neutral levels of AST and bilirubin. In addition, patients in this group tend to be older (average age 60.2), and group 6 has the second highest proportion of people with HIV (44. 4\%) and of these individuals, 75\% have advanced HIV. 

Another defining characteristic of clusters $5$ and $6$ is their shape. Despite our assumption of non-diagonal covariance, Figure \ref{fig:influential-pairs} shows that BUN and creatinine are highly correlated within clusters $5$ and $6$ (with a values of $0.69$ and $0.58$, respectively). In contrast, these values for the first four clusters are $0.40$, $0.44$, $0.24$, and $0.24$, respectively. Although this is a consequence of the inherent correlation between these variables, we can see that the magnitude of the correlation distinguishes the clusters.

Upon comparison, there are clinical characteristics shared by the clusters, particularly cluster 4 and cluster 6. Participants in these clusters tend to have neutral AST and elevated levels of BUN and creatinine. Despite these similarities in terms of the MR specifications, there are other factors that differentiate the subgroups. The average age of cluster 4 is 48.5, more than a decade younger than that of cluster 6, which is 60.2. Furthermore, Figure \ref{fig:influential-pairs} shows that, in general, the BUN and creatinine levels of group 6 are markedly higher than those of group 4. This distinction is picked up by the CLAMR model; the relatively low values of $\Delta_{j}^{4,*}$ for BUN and creatinine signal association of cluster 4 with several MRs. These differences may be useful for developing customized treatments for these clusters. In the general case, similar clusters can be combined after a post hoc clinical analysis or using a Gaussian kernel merging algorithm, e.g. the FOLD post processing method \citep{dombowsky2024bayesian}.

In light of these differential characteristics, we can summarize the interpretation of the clusters as follows. Cluster 1 comprises young female patients with generally neutral markers, low bilirubin (not a clinical concern), and low incidence of mortality; cluster 2 has modestly elevated levels of BUN that suggest possible renal dysfunction and skews older and more toward male sex than cluster 1; cluster 3 is characterized by increased frequency of HIV and elevated AST levels indicative of liver dysfunction; cluster 4 has elevated levels of BUN and creatinine that suggest septic shock or renal dysfunction; cluster 5 is composed mainly of men with low HIV prevalence/frequency and extremely high levels of AST indicative of liver dysfunction; and cluster 6 is composed of older individuals with extremely elevated BUN and creatinine indicative of septic shock or renal dysfunction.

\subsection{Comparison to SENECA Clusters}

Next, we qualitatively compare and contrast the INDITe clusters $\bs c^*$ with the four clusters reported in the SENECA study \citep{seymour2019derivation}, denoted as $\alpha$, $\beta$, $\gamma$, and $\delta$. Cluster 1 is comparable to their $\alpha$ cluster due to the lack of organ dysfunction and generally low mortality, and cluster 6 closely resembles their $\beta$-cluster as it is characterized by advanced age and renal dysfunction. Their $\delta$ cluster is associated with liver dysfunction; the elevated levels of AST in clusters 3 and, especially, 5 mean that liver dysfunction is also a defining factor in these INDITe clusters. However, we also find that factors specific to sSA explain the heterogeneity between the INDITe clusters. For example, malaria prevalence and low HIV prevalence are key characteristics of cluster 5, while cluster 3 is in part driven by low malaria prevalence and high HIV prevalence. Although cluster 1 is similar to the cluster $\alpha$, there are notable differences in demographics, including average age (37 years for INDITe, 60 years for SENECA) and sex (69\% female for INDITe, 49\% female for SENECA). However, we may interpret groups 1 and 6 as clinical manifestations of the groups $\alpha$ and $\beta$, respectively, in the northern Tanzanian population.  Regarding participant outcomes, we observe comparable, but distinct associations with those of the SENECA study, who found that the $\delta$ group (characterized by liver dysfunction) had the highest mortality rate in their cohort. We find that cluster 3 (characterized by HIV incidence and possible liver dysfunction) has the highest observed mortality at 18. 2\%, although cluster 4 (characterized by renal dysfunction) has a similar mortality at 17.9\%. 

\subsection{Comparison to Standard Methods}

\begin{figure}
    \centering
    \includegraphics[scale=0.6]{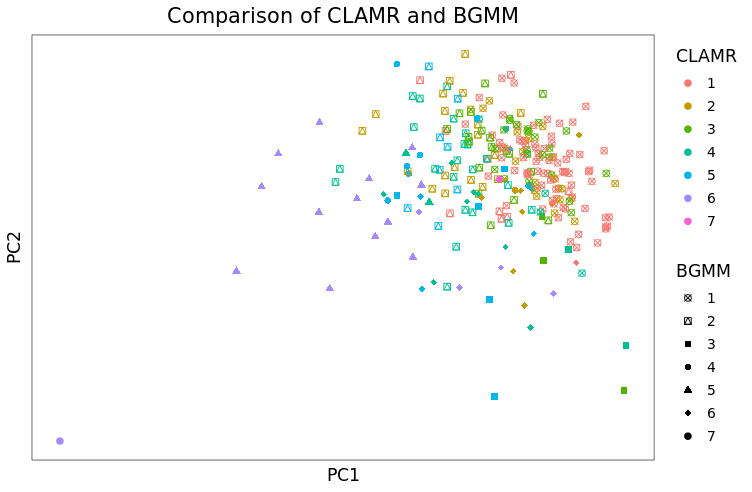}
    \caption{A comparison between the point estimate yielded by CLAMR and that by a standard BGMM. Colors indicate CLAMR, whereas shapes refer to the BGMM. Singleton clusters have been included.}
    \label{fig:CLAMR-vs-BGMM}
\end{figure}

First, we compare our results with the clustering point estimate obtained after fitting a Bayesian GMM (BGMM), as in \eqref{eq:yij-likelihood}, with weakly informative Gaussian and inverse-Gamma priors. Explicit details on hyperparameters are given in the Supplement. The adjusted Rand index between $\bs c^*$, the CLAMR point estimate and $\bs c^\prime$, the BGMM point estimate is $0.273$, indicating a modest resemblance between the two clusterings. Both $\bs c^*$ and $\bs c^\prime$ have 7 clusters and create singleton clusters (though not of the same participant). In contrast to the cluster sizes shown in Table \ref{table:demographic-information}, the sizes of the BGMM clusters are 125, 77, 26, 15, 11, 10, and 1. An explicit comparison between the two point estimates is given in Figure \ref{fig:CLAMR-vs-BGMM}, where it can be seen that the CLAMR clusters are split and subsequently merged to create those from the BGMM. For example, cluster 1 in CLAMR is spread over three clusters in the BGMM, and cluster 1 of the BGMM is comprised of participants from clusters 1, 2, 3, and 4 in CLAMR. The clusterings differ in interpretation as well. Clusters 3 and 4 of the BGMM, for instance, have an extremely small HIV incidence (10\% and 9.1\%), whereas the HIV incidence for all CLAMR clusters are larger than these proportions. Malaria incidences are generally higher in the BGMM clustering, with three of the seven clusters observed to have more than 20\% malaria incidence. All CLAMR-derived clusters have a lower incidence of malaria. Finally, BGMM places 202 of the individuals in just two clusters (76\% of the cohort), while CLAMR clusters more evenly divide the data set, making it easier to provide definitive interpretations. For these reasons, we find the results of CLAMR to be preferable from a clinical perspective. We further evaluate our methodology in the Supplemental by comparing the results from the CLAMR model to a Latent Class Model (LCM) \citep{linzer2011polca}. The latter model is implemented by dichotomizing the observed feature values based on which MR they are observed to be in.

\begin{figure}
    \centering
    \includegraphics[scale=0.65]{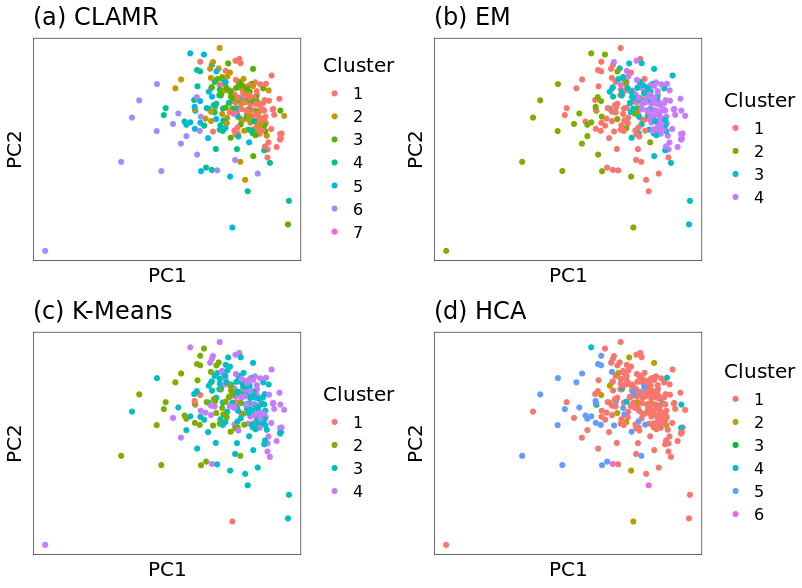}
    \caption{Comparison of the clusterings derived by EM, k-means, and complete linkage HCA on the INDITe data to the CLAMR point estimate. Colors correspond to cluster labels. Singleton clusters have been included.}
    \label{fig:standard-methods}
\end{figure}

Furthermore, we compare the results from CLAMR to several easily implementable clustering algorithms: the EM algorithm for a GMM, k-means, and complete linkage HCA. The GMM model used in the EM algorithm is determined by minimizing the Bayesian information criterion (BIC) for a selection of different covariance structures and number of components. The number of clusters in k-means is chosen by an elbow plot of the total within-cluster sum of squares, and we fix the number of clusters in HCA to be $6$, the same number of clusters in the CLAMR point estimate (with the singleton removed). The adjusted Rand indices between $\bs c^*$ and the clusterings of EM, k-means, and HCA are $0.385$, $0.042$, and $0.087$, respectively. A visual comparison between these methods is given in Figure \ref{fig:standard-methods}. There are some general patterns that emerge, such as a tendency for methods to cluster points with high values when projecting onto the first two principal components (i.e., in the upper right quadrant of the plots in Figure \ref{fig:standard-methods}), as well as points with lower values when projecting onto the first principal component (i.e., the middle sections of the plots). It is between these two extremes that the methods seem to differ, and this region is where clusters $2$-$5$ occur for the CLAMR clustering, $\bs c^*$. The differences in cluster definition for all methods in this region demonstrate the impact of using prior information when the data are not well-separated.

\subsection{Uncertainty Quantification}

The PSM for the INDITe data is shown in Figure \ref{fig:psm}, where each box along the diagonal indicates a cluster. Observe that both clusters 5 and 6 are notably distinct from cluster 1, i.e. $\Pr(c_i = c_j \mid \bs y)$ is very close to $0$ for $i \in C_1^*$ and $j \in (C_5^* \cup C_6^*)$. In fact, these clusters appear to be well separated among the influential variables in Figure \ref{fig:influential-pairs}, suggesting differing underlying subtypes and possibly differing treatment regimes. The PSM also indicates similarity between clusters 1, 2, 3, and 4 via the off-diagonal boxes in Figure \ref{fig:psm}. Comparing again to the within-cluster distributions in Figure \ref{fig:influential-pairs}, we can see that these clusters have similar distributions for the septic shock and renal dysfunction markers (BUN and creatinine). Hence, one would expect that the underlying subtypes manifest themselves similarly in the patient population, although differences in key immunologic factors such as HIV status and age, as well as AST and bilirubin levels, still distinguish these clusters in a clinical sense. Instead, it may be the case that individuals can embody characteristics of multiple subtypes. In the Supplement, we display histograms of the posterior distribution of $\Delta_{j}^{m,s_j^{m,*}}$ to convey uncertainty in our interpretative quantities; this is the posterior for $\Delta_j^{(m,k)}$ for $k=s_j^{m,*}$. These plots shed further light on our interpretations of the results. For example, the posteriors of $\Delta_{j}^{m,s_j^{m,*}}$ for cluster $4$ exhibit substantial uncertainty in MR association, whereas the posteriors for clusters 1 and 6 are more diffuse. We can also see multimodality for this quantity for creatinine in cluster 5; this is likely due to these participants being close to the border of the MRs.

\subsection{Posterior Predictive Samples and Sensitivity Analysis} \label{sect:posterior-predictive}
\begin{figure}
    \centering
    \includegraphics[scale=0.55]{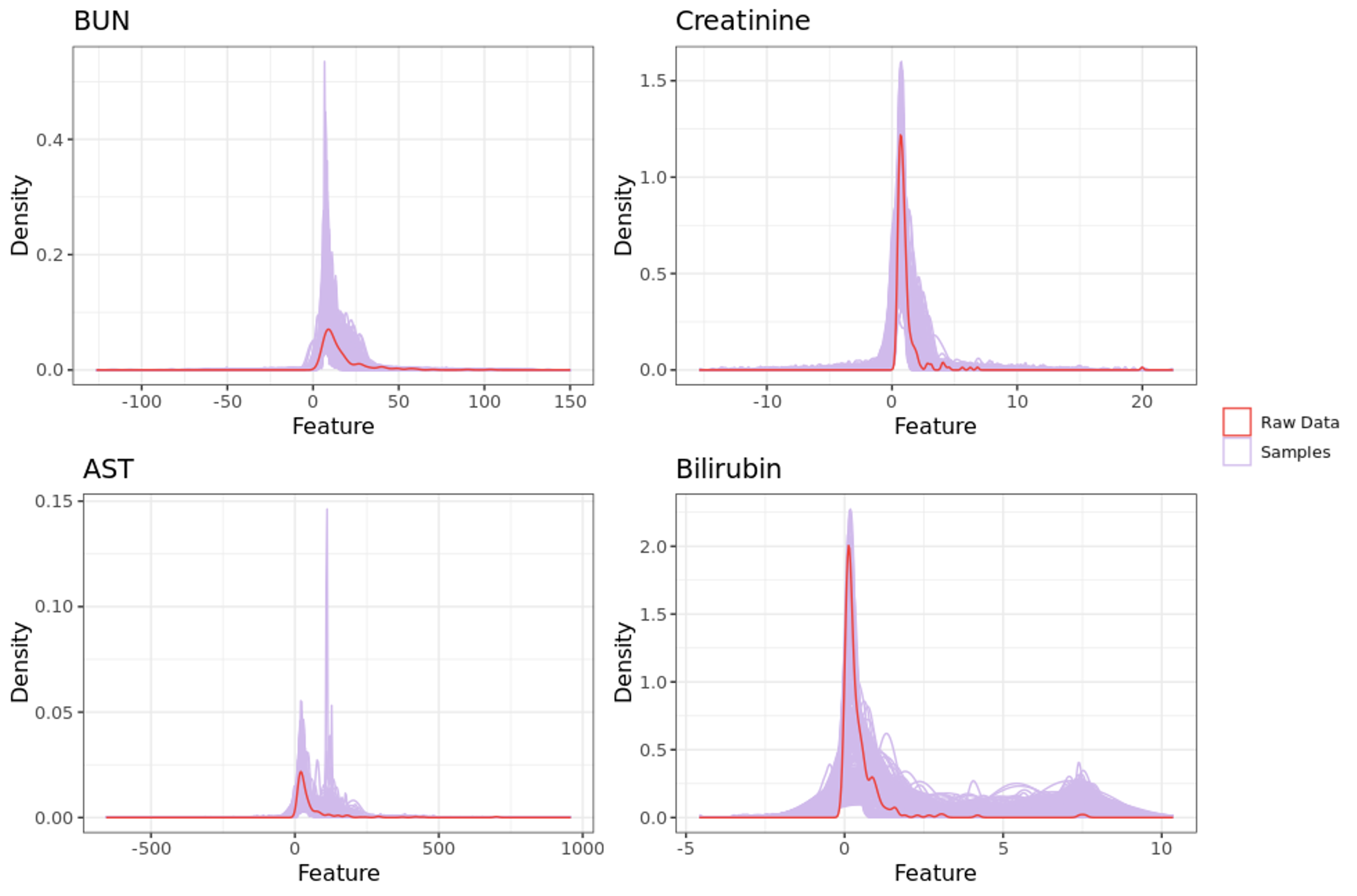}
    \caption{Summary of posterior predictive samples for the INDITe participants. The red line indicates the density of the data, whereas the violet curves are densities from $500$ posterior predictive samples. For visualization, missing features are imputed using multivariate imputation \citep{van2011mice}.}
    \label{fig:posterior-predictive}
\end{figure}

To gain insight into the capability of the CLAMR model to recapture patterns in the data, we sample from the posterior predictive distribution. For every measurement $y_{ij}$, we generate a posterior predictive sample $y_{ij}^{(t)}$ from \eqref{eq:yij-likelihood} using the Gibbs samples of the cluster centers, scales, and labels. The density functions for $500$ equally spaced posterior predictive samples across all chains are shown in Figure \ref{fig:posterior-predictive}, with the true density of the data displayed in red. Here, we can see that the CLAMR model is most accurate in predicting measurements below the 95\% percentile for all feature values. For measurements above the 95th percentile, the CLAMR model is notably less accurate at prediction, with a general tendency to cluster together measurements that are far in the tails of the empirical distribution. This behavior is explained by the MR specifications: since many of these tail measurements fall in the same MR, they are likely to cluster together. For example, measurements in the right tail for AST are all above the cut-off point for the second MR, and the CLAMR model rewards co-clustering these observations. This results in a mode in the posterior predictive samples midway through the right tail.

Figure \ref{fig:posterior-predictive} also highlights the differences between CLAMR and the standard BGMM. Generally, Bayesian Gaussian mixtures can excel in density estimation \citep{escobar1995bayesian}, although the interpretations of the resulting clusters may not be relevant to the application of interest. The specification of the CLAMR prior is motivated specifically by clustering and not density estimation, which means that the modes in the posterior predictive samples correspond to subgroups relevant to our clinical application and not necessarily actual Gaussian-shaped modes in the empirical distribution.

In the supplement, we apply CLAMR to the INDITe data with different MR specifications based on the SENECA results. These MRs are derived by translating the characteristics of the SENECA clusters to values relevant to the INDITe cohort (e.g., defining what expression levels for the variable $j$ characterize the clusters $\alpha$, $\beta$, $\gamma$ and $\delta$. Unlike the expression level MRs presented above, these MR specifications were allowed to be identical. There are differences and similarities in the results of both approaches. The SENECA MR specifications lead to more influential variables, but the expression level MRs lead to more clusters. However, BUN, creatinine, bilirubin, and AST clusters are derived using both MR specifications, which means that there are some common clinical factors driving the results, regardless of the MRs chosen.

\section{Discussion} \label{sect:discussion}

In this article, we present CLAMR, a novel enhancement of the Bayesian Gaussian mixture model motivated by disease subtype analysis. The main idea behind our approach is to incorporate commonly used cut-offs in feature values, or meaningful regions, into the prior for the cluster centers, with the ultimate goal being to distinguish clusters via their medical interpretations. We accomplish this mathematically by assuming that the prior on the cluster centers is a mixture model over the MRs. When the CLAMR prior is a Gaussian mixture, a simple Gibbs sampler can be derived for posterior computation. The algorithm performs excellently on synthetic data in which the Gaussian model is misspecifed. In addition to clustering, our prior can be used for a pre-training procedure to select relevant clustering features using the results of Bayesian hypothesis tests.

CLAMR is primarily motivated by the clustering challenges presented by the INDITe sepsis data from northern Tanzania, such as the lack of separation in the data and the small sample size. We find that the influential features are all markers of organ dysfunction and, using these, we derive $6$ clusters of varying size and interpretation. These groups are related to clinical factors and signs, including HIV prevalence, malaria status, age, sex, and kidney and liver dysfunction, and we also observe varying mortality rates across the groups. In addition, several of our clusters are similar in interpretation to the clusters of the SENECA study, despite the fact that the sepsis patient population in sSA has markedly different distributions of age, comorbidities and etiologies than those in North America.

There are multiple possible extensions to CLAMR that can generalize our approach. The cluster profiles $S_j$ can be estimated after applying label-switching algorithms to the samples $ \c^{(t)}$ and $\s^{(t)}$, then minimizing a clustering loss such as Binder's or VI. The CLAMR prior is presented here as a GMM, but any continuous probability distribution can be used to model the prior density within the MRs. If investigators require the distribution within MRs to be multimodal, for example, one can set the component kernels of $\pi(\mujl)$ to be themselves GMMs, and posterior computation can be carried out using a similar Gibbs sampler to the one we utilize. We assume that the cluster-specific variances follow weakly-informative inverse-gamma priors, but this can be simply augmented to a mixture of inverse-gamma kernels over the MRs in order to model varying shapes, not just centers, between profiles. Finally, in this article, we implement the CLAMR prior for cluster centers in a GMM with diagonal covariance matrices. The CLAMR prior could also be applied to a GMM with non-diagonal covariance, and we provide analytic details of this extension in the supplementary material.

Our framework is a two-step procedure: first, we screen for influential variables, then report the clustering results of the reduced dataset. We perform these tasks in two separate steps to (a) formalize the choice of influential features as a hypothesis testing problem, and (b) to avoid the negative theoretic properties of Bayesian clustering that arise even in a modest dimension \citep{chandra2023escaping}. However, CLAMR could be extended to a one-step procedure, in which an appropriate variable selection prior is used to reduce the feature set. For example, we could incorporate latent discrimination vectors $\alpha_j \in \lb 0, 1\rb$ for all characteristics $j = 1, \dots, p$ \citep{tadesse2005bayesian, kim2006variable}. When $\alpha_j=1$, we include this feature in the GMM likelihood and draw $j$ from the CLAMR prior; if $\alpha_j=0$, feature $j$ is modeled with a Gaussian distribution with conjugate Gaussian-inverse-gamma priors.

In addition to the MRs, other sources of prior information can be incorporated into the CLAMR algorithm. Our current approach relies on overfitting the GMM, but if a reliable source on the number of clusters is available (either from expert information or from a pre-training step using the EM algorithm), this can be used by either (a) fixing $L$ to be this estimate or (b) using this estimate to choose the hyperparameters in a clustering loss function. If, instead, clinicians have an informed guess of the clustering, one could use a centered partition process \citep{paganin2021centered} for the prior on $\bs c$. The CLAMR prior can also be extended to more complicated regimes that frequently arise in clinical analyses, such as multiview clustering \citep{franzolini2023bayesian}. 

\section*{Acknowledgments}
This work is supported in part by funds from the National Institutes of Health under grant numbers R01AI12137, R01AI155733, and R01ES035625; the Office of Naval Research under grant number N00014-21-1-2510; and the National Institute of Environmental Health Sciences under grant number R01ES027498. Dombowsky is also funded by the Myra and William Waldo Boone Fellowship.

\bibliographystyle{apalike}
\bibliography{sources}

\section*{Supplementary Material}
The supplementary material includes: (a) Extended computational details including a Gibbs sampler; (b) A description of alternate notions of MRs; (c) Technical details including hyperparameter choices for the simulation study; (d) a simulation study for a well-specified Gaussian mixture; (e) Hyperparameter choices, MCMC diagnostics, and additional figures for the application to the INDITe data; (f) A comparison of our results to a latent class model (LCM); (g) An additional analysis of the INDITe data using MRs derived from the patient clusters described by SENECA; (h) An extension of the CLAMR prior for categorical data; and (i) Technical details on extending CLAMR to non-diagonal covariance matricies. Code implementing the simulation studies can be found at \url{https://github.com/adombowsky/CLAMR}.

\appendix

\section{Gibbs Sampler}
Posterior computation with the CLAMR prior requires only a few modifications to existing Gibbs samplers for finite GMMs, e.g. the algorithm in \cite{maclachlanfinite}. 
We must sample from the full conditional distriubtions of the clustering $\c$, the cluster-specific parameters $\lb (\mujl, \sigmajlsq) \rb_{l=1}^L$, and the profiles $\s$. Note that we have omitted the random probability vectors $\bs \psi$ and $\lb \bs \phi_j \rb_{j=1}^p$ so as to improve MCMC mixing, especially when sampling the multinomial groupings $\c$ and $\s_j$. 
We update the cluster allocations via $\pi(c_i = l \mid -) \propto \lb \prod_{j=1}^p \N(y_{ij};\mujl, \sigmajlsq) \rb \lb n_l^{(-i)} + \gamma/L \rb;$
for $l=1, \dots, L$, where $n_l^{(-i)} = \sum_{i^\prime \neq i} \textbf{1}(c_{i^\prime} = l)$. A similar identity holds for reallocation of the $l$th label in feature $j$, i.e. $\pi(s_j^{(l)} = k \mid -) \propto \N(\mujl; \xijk, \taujksq) \lb m_k^{(-l)} + \rho_j/K_j \rb$, with $m_k^{-(l)} = \sum_{l^\prime \neq l} \textbf{1}(s_j^{(l^\prime)} = k)$. For the cluster centers $\mujl$, we sample from an updated Gaussian distribution, which depends on the data allocated to cluster $l$ and the hyperparameters of meaningful region $s_j^{(l)}$,
\begin{equation*}
    \pi(\mujl \mid s_j^{(l)} = k, -) \propto \N \left(\mujl;   \frac{ \frac{\xijk}{\taujksq} + \frac{\sum_{c_i=l} y_{ij}}{\sigmajlsq}}{\frac{1}{\taujksq} + \frac{n_l}{\sigmajlsq}}, \left( \frac{1}{\taujksq} + \frac{n_l}{\sigmajlsq} \right)^{-1}   \right),
\end{equation*}
where $n_l = \sum_{i=1}^n \textbf{1}(c_i=l)$. The full conditional distribution of $1/\sigmajlsq$ is the standard updated Gamma distribution, $\tx{G}(\lambda_j + n_l/2, \beta_j + \sum_{i:c_i=l} (y_{ij} - \mujl)^2/2)$. Any missing values are imputed each step of the Gibbs sampler by simulating a draw from the Gaussian likelihood using the current values of $c_i$, $\mujl$, and $\sigmajlsq$.

\section{Alternate Notions of Meaningful Regions}
Level of expression may not be the only way in which MRs arise in applications. If a previous study has clustered a similar cohort 
into $K$ groups using the same $p$ variables, then $\M_j^{(k)}$ can represent the support of variable $j$ in previously observed cluster $k$. In our application, the Sepsis Endotyping in Emergency Care (SENECA) study clustered sepsis patients from $12$ hospitals in Pennsylvania, USA using the same clinical variables and provides summaries of the distribution of data within each cluster \citep{seymour2019derivation}. The event $y_{ij} \in \M_j^{(k)}$ can be interpreted as individual $i$ exhibiting characteristics emblematic of previously observed cluster $k$. Alternatively, MRs may correspond to risk scores that predict mortality. Examples from our population of interest include the modified early warnings score (MEWS) \citep{subbe2001validation} and the universal vital assessment (UVA) score \citep{moore2017derivation}. Both the MEWS and UVA score are calculated by determining the total ``points" an individual receives based on cut-offs of the features. We can then set $\M_j^{(k)}$ to be the set cut-offs for variable $j$ that correspond to different point values. In the main article we focus on the case of expression level MRs, but alternate strategies for other kinds of MRs may be useful in other settings. 

\section{Additional Details on Simulations}
First, we detail technical specifications for the simulation study in Section \ref{sect:sims-misspec}. The true MRs, which are assumed to be known, are given in Table \ref{table:true-MR-simulations}. We simulate the centers means $\tilde \mu_{j}^{(l)}$ from the CLAMR prior with our default hyperparameters and $\omega = 0.99$, encouraging cluster centers that are near the midpoint of each interval. The scales $\tilde \sigma_j^{(l)}$ are selected so that simulated data tends to concentrate in each true MR. For true clusters $1$ and $2$, the true scales across the six features are $(1/6,1/2,0.6,0.05,20,20)$, and for true cluster $3$, these are $(1/6,1/2,1/3,0.05,20,20)$ (i.e., we simulate slightly different kernel shape in the third cluster). In each replication, we run the CLAMR and BGMM Gibbs samplers for $15,000$ iterations each, with the first $5,000$ discarded as burn-in. For CLAMR, the variance priors are $1/\sigmajlsq \sim \tx{G}(10/R_j, 10)$, where $R_j = b_j^{K_j} - a_j^{(1)}$ is the general range of feature $j$. For the BGMM, we center and scale the synthetic data to have mean $0$ and standard deviation $1$. The priors for the BGMM are $\mu_j^{(l)} \sim \N(0,1)$ and $1/\sigma_j^{(l)2} \sim \tx{G}(1,1)$. The point estimates for both Bayesian methods are computed by minimizing the VI loss. The EM algorithm is implemented in the package \texttt{mclust} \citep{scrucca2016mclust}, where the number of clusters is selected by minimizing the Bayesian information criterion over a set of candidate models. We fix the maximum number of clusters in the EM algorithm to be $5$. For both k-means and HCA, the number of clusters is fixed at the truth (i.e., $3$). 
\begin{table}[ht]
\centering
\begin{tabular}{rrrrrrr}
  \hline
 Feature & 1 & 2 & 3 & 4 & 5 & 6 \\ 
  \hline
D & [-1,1] & [0,2] & [-10,2] & [-0.1,0.1] & [0,100] & [0,10] \\ 
  N & [1,2] & [2,5] & [2,4] & [0.1,0.3] & [100,250] & [10,30] \\ 
  E & [2,4] & [5,10] & [4,8] & [0.3,0.5] & [250,400] & [30,200] \\ 
   \hline
\end{tabular}
\caption{The true MRs for the misspecified simulation study, ordered by their level of expression (D, N, and E).}
\label{table:true-MR-simulations}
\end{table}

The simulations in Section \ref{sect:sims-no-MRs} are generated in the same manner, with the exception that the true centers no longer are simulated from the CLAMR prior. Instead, we set $ \tilde{\mu}_{j}^{(l)} \sim \N(\xi_j, \tau_j^2)$, where $(\xi_j, \tau_j)$ are chosen by applying our default hyperparameters for a \textit{single} MR with endpoints $a_j^{(1)}$ and $b_j^{(K_j)}$, and density parameter $\omega=0.95$. CLAMR hyperparameters, point estimation, and competitors are also identical to the simulations in Section \ref{sect:sims-misspec}. 

\section{Well-Specified Simulation Study}

\begin{table}[ht]
\centering
\begin{tabular}{rrrrrrr}
  \hline
 Feature & 1 & 2 & 3 & 4 & 5 & 6 \\ 
  \hline
D & [-1,10] & [0,10] & [-10,2] & [-0.1,0.1] & [0,100] & [0,10] \\ 
  N & [10,20] & [10,25] & [2,4] & [0.1,0.3] & [100,250] & [10,30] \\ 
  E & [20,40] & [25,50] & [4,8] & [0.3,0.5] & [250,400] & [30,200] \\ 
   \hline
\end{tabular}
\caption{The true MRs for the well-specified simulation study, ordered by their level of expression (D, N, and E).}
\label{table:wellspec-true-MR-simulations}
\end{table}

\begin{table}[ht]
\centering
\begin{tabular}{rrrrrrr}
  \toprule
 & $n$ & CLAMR & BGMM & EM & k-means & HCA \\ 
  \midrule
ARI & $100$ & 0.99 (0.06) & 0.94 (0.14) & 1.00 (0) & 0.76 (0.24) & 0.80 (0.23) \\
  & $500$ & 0.99 (0.08) & 0.97 (0.11) & 1.00 (0) & 0.79 (0.23) & 0.78 (0.22) \\ 
  & $750$ & 1.00 (0.00) & 0.99 (0.07) & 1.00 (0) & 0.79 (0.23) & 0.79 (0.22) \\ 
  & $1000$ & 0.98 (0.08) & 0.99 (0.06) & 1.00 (0) & 0.78 (0.24) & 0.76 (0.25) \\ 
   \midrule
 $\hat{L}$ & $100$ & 2.98 (0.14) & 2.86 (0.35) & 3.00 (0) & 3.00 (0) & 3.00 (0) \\ 
  & $500$ & 2.97 (0.17) & 2.94 (0.24) & 3.00 (0) & 3.00 (0) & 3.00 (0) \\
  & $750$ & 3.00 (0.00) & 2.97 (0.17) & 3.00 (0) & 3.00 (0) & 3.00 (0) \\ 
  & $1000$ & 2.96 (0.20) & 2.98 (0.14) & 3.00 (0) & 3.00 (0) & 3.00 (0) \\ 
\end{tabular}
\caption{ARI with $\tilde{\bs c}$ and $\hat L$ for CLAMR, the BGMM, EM, k-means, and HCA applied to $100$ independent replications for each $n \in \lb 100, 500, 750, 1000 \rb$. Both k-means and HCA have the number of clusters fixed at $L=3$, the true value. Standard deviations of the metrics across the replications are given in brackets. A value of $(0.00)$ for the standard deviation indicates that the unrounded value is larger than $0$.}
\label{table:well-specified-sims}
\end{table}

While mixture models are in general misspecified, we are interested in evaluating how the CLAMR prior impacts cluster inference when the data are simulated from a GMM. We perform an additional simulation where we generate $y_i \mid \tilde c_i = l \sim \N(\tilde \mu_j^{(l)}, \tilde \sigma_j^{(l)2})$. Unlike in Section \ref{sect:sims-misspec}, we simulate $\Tilde{\bs s}$ randomly for each replication, where $\Pr(\tilde s_j^{(l)} = k) = 1/3 $ for $k=1,2,3$. The true MRs are given in Table \ref{table:wellspec-true-MR-simulations}. Otherwise, the simulation settings (e.g., the true scales $\tilde \sigma_j^{(l)}$ and generation of cluster centers) are the same as in the simulations in Section \ref{sect:sims-misspec}. As before, we generate 100 independent replications of this experiment for each choice of sample size $n \in \lb 100, 500,750,1000 \rb$. The results are displayed in Table \ref{table:well-specified-sims}. As expected, the EM algorithm excels, whereas CLAMR and BGMM occasionally underestimate the true number of clusters, resulting in a lower average ARI. However, we can see that CLAMR outperforms the BGMM for $n \leq 750$, with the BGMM having slightly better performance for $n=1000$. In addition, we can see that all model-based methods outperform k-means and HCA, although the true number of clusters is known for these methods.

\section{Additional Details on Application to INDITe Data}

\subsection{Hyperparameters and MCMC Diagnostics}

In this section, we expand on the technical aspects of our main application. Our priors for cluster-specific variances are $1/\sigmajlsq \sim \tx{G}(2, 1/R_j)$, where $R_j = b_j^{K_j} - a_j^{(1)}$ is the range of feature $j$. This prior is essentially a limiting case of the hierarchical variance prior used in \cite{richardson1997bayesian} for Gaussian mixtures. Both the initial pre-training phase and the final clustering phase were fit using the same hyperparameters and number of iterations and chains. We fit the CLAMR model using $9$ independent chains in parallel, where each chain has $300,000$ iterations, with the initial $10,000$ removed as burn-in. The CLAMR Gibbs sampler took 2750.272 seconds (45.838 minutes) with the four influential variables as the feature set (i.e., after the pre-training phase). We then thin the MCMC output using only every $10$ iteration to compute the Bayes factors and the clustering point estimate.

\begin{figure}
    \centering
    \includegraphics[scale=0.6]{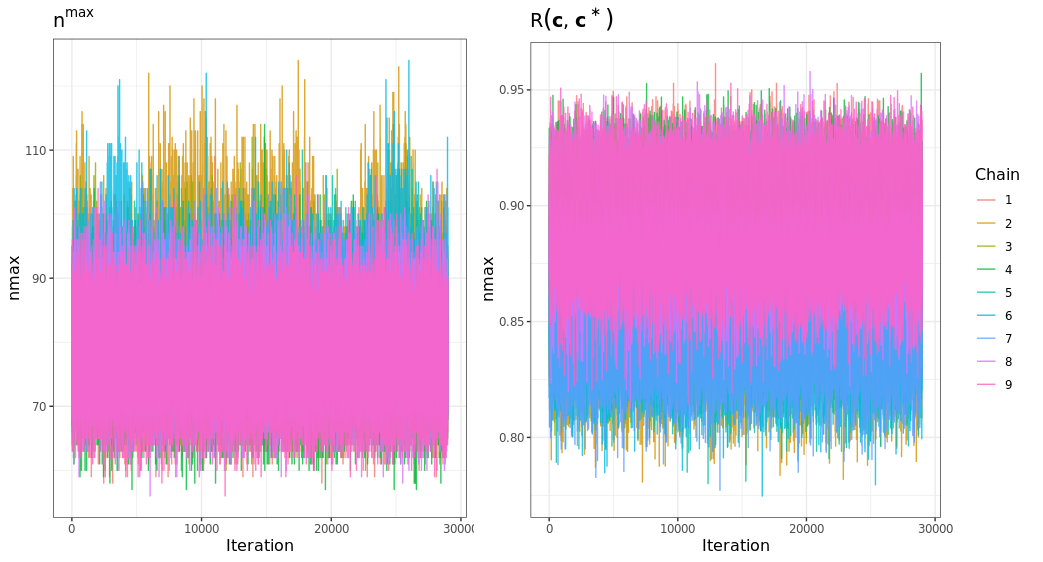}
    \caption{Traceplots of both the maximum cluster size (left) and Rand index with $\bs c^*$ (right) across the nine chains.}
    \label{fig:traceplots}
\end{figure}

The left plot in Figure \ref{fig:traceplots} shows trace plots for the maximum cluster size, that is,
\begin{equation*}
    n^{\max} = \max(|C_1|, \dots, |C_{L}|).
\end{equation*}
The value of $\hat R$ \citep{gelman1992inference} for $n^{\max}$ is 1.070.  We further evaluate the MCMC chains of the partition posterior by quantifying their clustering point estimation properties. First, we compute separate clustering point estimates for each chain and then calculate the Rand index \citep{rand1971objective} between each pair of point estimates. The lowest pairwise Rand index is 0.88, the highest is 1.00, and the average across all chains is 0.95, indicating substantial similarity between the point estimates. Second, we compute the Rand index between the partition samples and $\bs c^*$, denoted $R(\bs c, \bs c^*)$; the trace plots of this quantity are displayed in the right plot of Figure \ref{fig:traceplots}. Here, we can see that all chains are exploring partitions similar to $\bs c^*$, although there is some variation across chains.

Regarding the cluster-specific parameters, we obtain MCMC samples of the log-likelihood conditional on the cluster labels, $\log  \mathcal L(\bs y; \bs c, \bs \mu, \bs \sigma)$, which attains an $\hat R$ value of $1.738$. This suggestion of inadequate convergence or mixing is likely driven by the sampling behavior of the cluster centers $\mujl$ due to the multimodality of the prior and posterior. Hence, we state all of our inferences in terms of the posterior of $\bs c$, which we find to have adequate mixing based on the monitoring of simple summary statistics of the partitions. We do find that the posterior predictive samples do not show clear evidence of a lack of fit of the model to the data.

The BGMM is implemented by first centering and scaling the original $p=15$ variables to have mean $0$ and unit variance. A priori, we assume $\mu_j^{(l)} \sim \N(0,1)$ and $1/\sigma_j^{(l)2} \sim \tx{G}(1,1)$. The number of components is set to $L=10$, and the cluster probabilities $\bs \psi \sim \tx{Dir}(1/L, \dots, 1/L)$. We run a Gibbs sampler with $200,000$ iterations with the first $10,000$ discarded as burn-in. Point estimates are computed by minimizing the VI loss.

\subsection{Additional Figures}
\begin{figure}
    \centering
    \includegraphics[scale=1.1]{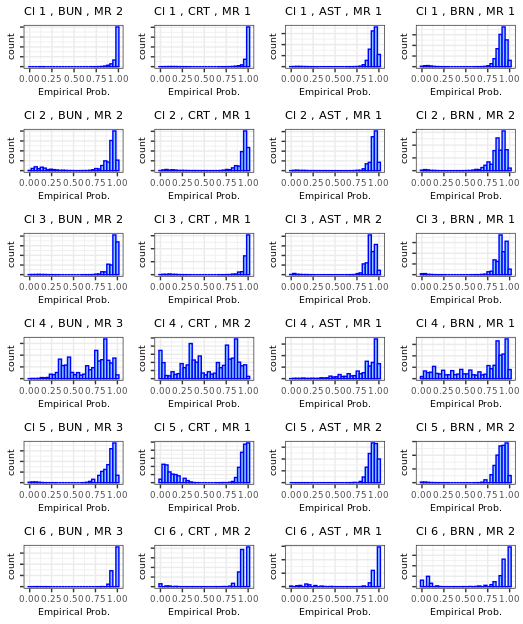}
    \caption{Histograms for the posterior samples of $\Delta_j^{m,k}$, with $k=s_j^{m,*}$, across the six clusters and four influential variables. Here, ``Cl" abbreviates ``Clusters", ``CRT" abbreviates ``creatinine", and ``BRN" abbreviates ``bilirubin".}
    \label{fig:delta-histograms}
\end{figure}

\begin{figure}
    \centering
    \includegraphics[scale=0.5]{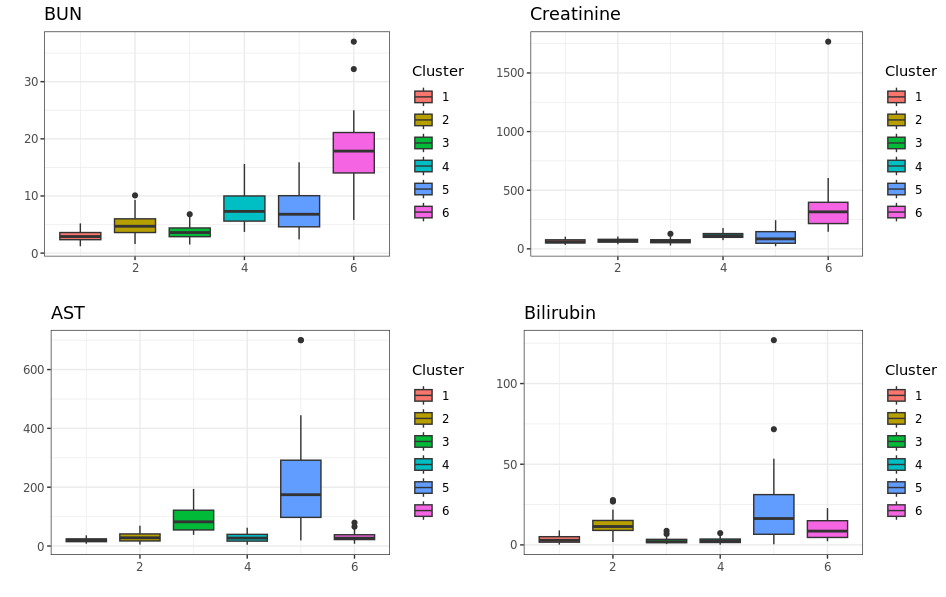}
    \caption{Boxplots showing the within-cluster distribution of the influential features.}
    \label{fig:box-plots}
\end{figure}

\begin{figure}
    \centering
    \includegraphics[scale=0.65]{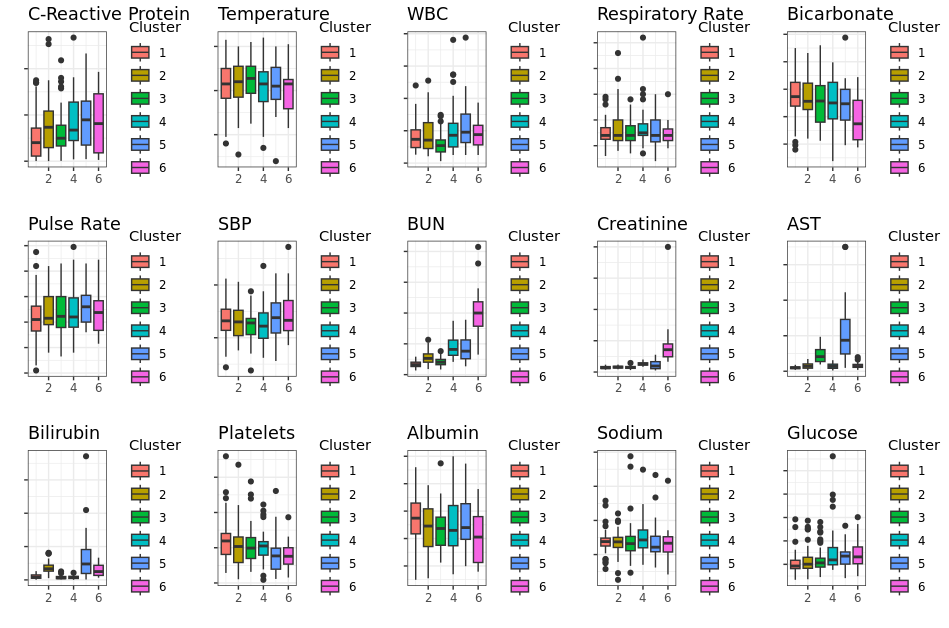}
    \caption{Boxplots showing the within-cluster distribution across all original 15 features.}
    \label{fig:full-boxplots}
\end{figure}

We display boxplots for the within-cluster distributions across the influential features in Figure \ref{fig:box-plots}. Here, we can see that the increased values in BUN and creatinine disinguish cluster 6 from the other groups, while the same holds for cluster 5 and the increased values of both AST and bilirubin. We also see relatively high values of AST in cluster 3 and relatively high values of bilirubin in cluster 2. Boxplots for all 15 original features are displayed in Figure \ref{fig:full-boxplots}. Histograms of the posterior distribution of $\Delta_j^{m,k}$ with $k=s_j^{m,*}$ are shown for the six clusters in Figure \ref{fig:delta-histograms}.

\subsection{Comparison to a Latent Class Model (LCM)}

\begin{table}
\centering
\begin{tabular}{llllllll}
\toprule
   & Overall & $C_1^{\tx{LCM}}$   & $C_2^{\tx{LCM}}$   & $C_3^{\tx{LCM}}$   & $C_4^{\tx{LCM}}$   & $C_5^{\tx{LCM}}$ & $C_6^{\tx{LCM}}$    \\
\midrule
Size    & 264 & 26   & 25    & 69    & 26    & 79   & 39     \\
AST & * & N & N & N & N & N & N \\
Bilirubin & * & N & N & D & N & D & N \\
BUN & *  & E & N & N & N & N & O \\
Creatinine & * & E & N & N & N & N & E\\
\midrule
\end{tabular}
\caption{MR interpretations for the LCM applied to $\bs x$, where each entry is computing by maximizing the posterior MR probability within each class. The influential variables from the analysis in the main article are displayed. The categories are D, N, E, and O, the latter of which stands for ``outlier" (i.e., a right tail observation for the final MR).}
\label{table:LCM-info}
\end{table}

In addition to standard clustering methods for continuous features, we apply a Latent Class Model (LCM) to the INDITe participant data. First, we impute missing values using the R package \texttt{mice} \citep{van2011mice}, which we also use to initialize missing data imputation in CLAMR. Then, we coarsen the INDITe features $\bs y$ into discrete data $\bs x$ via the map $x_{ij} = k \iff y_{ij} \in [a_j^{(k)}, b_j^{(k)}]$. For any notable outliers (that is, measurements with $y_{ij} > b_j^{(K_j)}$), we set $x_{ij} = K_j+1$. We then apply the \texttt{poLCA} package \citep{linzer2011polca} to the coarsened data $\bs x$. For level comparison with the CLAMR point estimate $\bs c^*$, we fix the number of classes in \texttt{poLCA} to be $6$. The algorithm is run for a maximum of $10,000$ iterations. We denote the vector of predicted class memberships for the INDITe participants by $\bs c^{\tx{LCM}}$.

The adjusted Rand index between $\bs c^*$ and $\bs c^{\tx{LCM}}$ is $0.213$, indicating substantial differences in subgroup membership between the two approaches. For a concise comparison, we will discuss the interpretations of the $\bs c^{\tx{LCM}}$ clusters with respect to the influential variables BUN, creatinine, AST, and bilirubin. To derive MR interpretations for the different classes, we maximize the posterior probability of MR membership. These results are displayed in Table \ref{table:LCM-info}. First, we can see that the class sizes differ from those in $\bs c^*$, which is suggested by the low adjusted Rand index between the two groupings. Furthermore, AST does not distinguish the classes from each other, as each class is associated with neutral feature expression, though AST is a key variable for interpreting the CLAMR clusters. There are some common characteristics between the two clusterings despite their dissimilarity. For example, both $\bs c^*$ and $\bs c^{\tx{LCM}}$ have two subgroups associated with high levels of BUN and creatinine; both approaches also lead to large clusters with general neutral markers and diminished bilirubin. Finally, $\bs c^{\tx{LCM}}$ leads to some classes with similar MR associations, e.g., classes 2 and 4, and classes 3 and 5.

\section{Clustering the INDITe Data with SENECA Centered MRs}

\begin{table}[]
\centering
\begin{tabular}{lll}
\toprule
  Feature  & Type & $\tx{BF}_j$    \\
\midrule
Temperature         & \multirow{4}{7em}{Clinical Signs} & 5.257    \\
Respiratory Rate       &  & 0.948    \\
Pulse Rate         &    & 0.585    \\
Systolic Blood Pressure (SBP)  &    & 0.231    \\
\midrule
White Blood Cell Count (WBC)    & \multirow{2}{7em}{Inflammation Markers}   & $\bs \infty $   \\
C-Reactive Protein (CRP)  &  & \textbf{24.744}  \\
\midrule
Aspartate Transaminase (AST)         &  \multirow{5}{7em}{Organ Dysfunction Markers}   & \textbf{327020.149}   \\
Bilirubin         &    & \textbf{22.637}   \\
Blood Urea Nitrogen (BUN)  & &    $\bs \infty$  \\
Creatinine &    & \textbf{35.645}  \\
Platelets   &  & \textbf{31.040}  \\
\midrule
Glucose   &  \multirow{4}{7em}{Additional Markers} & 2.868    \\
Sodium     &  & 5.898 \\
Bicarbonate      &    & 1.258  \\
Albumin     &  & \textbf{65.789}  \\
\bottomrule
\end{tabular}
\caption{The complete 15 variables from the INDITe data with accompanying Bayes factors from the pre-training phase using the SENECA clusters as MRs. Here, $\tx{BF}_j = \infty$ indicates that $\pi(H_{0j} \mid \bs y) = 0$.}
\label{table:SENECA-features-and-bfs}
\end{table}

We now present an alternative implementation of CLAMR to the INDITe data using a different notion of MR based on the sepsis patient clusters derived by the SENECA study. Recall that the SENECA study obtained $4$ clusters--$\alpha$, $\beta$, $\gamma$, and $\delta$--from a cohort spread across twelve hospitals in Pennsylvania \citep{seymour2019derivation}. Our expert clinical collaborators translated the medical characteristics described in \cite{seymour2019derivation} to the northern Tanzania population, adjusting for key demographic differences including age and etiologies. Ultimately, our collaborators were able to provide information on the expression levels characteristic of each SENECA cluster, were they to manifest themselves in sSA. For example, BUN would be neutral (N) in the $\alpha$ cluster, elevated (E) in the $\beta$ cluster, neutral in the $\gamma$ cluster, and elevated in the $\delta$ cluster. Therefore, for BUN, we set the four MRs to be $\lb \tx{N}, \tx{E}, \tx{N}, \tx{E} \rb$.  Unlike our application in the main article, the MRs are \textit{not} disjoint, and we exclude any expression levels not represented by the SENECA clusters. For BUN, for instance, it is not possible for a cluster to have a profile that implies diminished (D) expression level, meaning that the prior support has been constrained.

\begin{table}
\centering
\begin{tabular}{llllll}
\toprule
   & Overall & $C_1^*$   & $C_2^*$   & $C_3^*$   & $C_4^*$     \\
\midrule
Size    & 265 & 107   & 105    & 35    & 18          \\
$\Pr$(Female) & 0.487  & 0.477 & 0.590 & 0.229 & 0.444     \\
$\Pr$(HIV+)  & 0.381   & 0.336 & 0.467 & 0.229 & 0.444    \\
$\Pr$(Advanced HIV) & 0.257  & 0.196 & 0.324 & 0.200 & 0.333  \\
$\Pr$(Malaria) & 0.098 & 0.168 & 0.143 & 0.143 & 0.167 \\
$\Pr$(Deceased) & 0.125 & 0.131 & 0.105 & 0.143 & 0.167    \\
Avg. Age (SD) & 42.8 (17) & 45.1 (18.3) & 38.8 (14) & 40.2 (15.8) & 57.3 (17.7)  \\
\bottomrule
\end{tabular}
\caption{Basic demographic information for the clusters derived using SENECA centered MRs.}
\label{table:SENECA-demographic-information}
\end{table}

We run 10 MCMC chains in parallel for $200,000$ iterations (with the first $10,000$ discarded as burn-in, then thinned to every tenth iteration) and all hyperparameters are identical to those for the INDITe application in the main article, with the exception that $K_j = 4$ for all features, and we set $\rho_j = 0.69$. Table \ref{table:SENECA-features-and-bfs} shows the Bayes factors for all features from the original dataset. Interestingly, the model selects a less sparse feature set, and includes all inflammation and organ dysfunction markers, as well as the additional marker albumin. After pre-training, the clustering point estimate is obtained, which we summarize in Table \ref{table:SENECA-demographic-information}. We find four clusters, with sizes 107, 105, 35, and 18. 

\begin{table}[ht!]
\centering
\begin{tabular}{rrrrrrr}
  \hline
 & SENECA & 1 & 2 & 3 & 4 \\ 
  \hline
& \multirow{5}{7em}{Expression Levels} 1 &  31 &  56 &   0 &   0 \\ 
 & 2 &  37 &  11 &   1 &   0 \\ 
 & 3 &   5 &  33 &   6 &   0 \\ 
 & 4 &  32 &   5 &   0 &   2 \\ 
 & 5 &   0 &   0 &  27 &   0 \\ 
 & 6 &   2 &   0 &   0 &  16 \\ 
 & 7 &   0 &   0 &   1 &   0 \\ 
   \hline
\end{tabular}
\caption{Contingency table between the clustering derived using expression level MRs (rows) and that using SENECA centered MRs (columns).}
\label{table:SENECA-continingency-table}
\end{table}

There are some similarities between the clusters in Table \ref{table:SENECA-demographic-information} and those derived using expression level MRs. In both analyses, there are clusters driven by increased values of BUN and creatinine as well as increased values of bilirubin and AST. In addition, one of the clusters is associated with older age (cluster 6 using expression level MRs, and cluster 4 using SENECA MRs). In terms of differences, the adjusted Rand index between the two point estimates is $0.260$, and we visualize the comparison between the two clusterings using the contingency table in Table \ref{table:SENECA-continingency-table}. Note the disparity in cluster sizes between the two analyses; the model using SENECA MRs prefers to allocate most of the data into two large clusters. Perhaps the biggest difference is that cluster 1 using expression level MRs is split between two clusters using SENECA MRs. In addition, the SENECA MR analysis does not include a singleton cluster, and it allocates that participant into cluster 3. However, we can see that there is significant overlap between clusters 4, 5, and 6 using expression level MRs and clusters 1, 3, and 4 using SENECA MRs. To compare the two analyses more concretely, we compute the Watanabe–Akaike information criterion (WAIC) \citep{watanabe2010asymptotic} for both models. The WAIC under expression level MRs is 7.826, whereas for SENECA MRs, it is 37.615.

\section{Extension to Categorical and Mixed Data}
We now discuss a generalization of our approach for categorical and binary features. In medical applications, it is common for a study to gather information related to comorbidities, etiologies, and presence-absence. These features are generally taken to be categorical, i.e. $y_{ij} \in \lb 1, \dots, M_j \rb$, where $1<M_j<\infty$ is the number of categories for the $j$th feature. We model the feature using a mixture of multinomial distributions, 
\begin{equation*}
    y_{ij} \mid c_i = l, \bs p_j^{(l)} \sim \tx{Multinomial}(\bs p_j^{(l)}),
\end{equation*}
where $\bs p_j^{(l)} = (p_{j1}^{(l)}, \dots, p_{jM_j}^{(l)})$, $0<p_{jm}^{(l)}<1$ and $\sum_{m=1}^{M_j} p_{jm}^{(l)}=1$. Here, we assume that each category in the $j$th feature has a medical interpretation, meaning that the categories $\lb 1, \dots, M_j \rb$ are the MRs. Similar to our original formulation for a Gaussian mixture, we specify a prior that is a mixture of informative distributions for each MR,
\begin{gather} \label{eq:multinomial-CLAMR}
    \bs p_j^{(l)} \sim \sum_{m=1}^{M_j} \phi_j^{(m)} \tx{Dir}( \nu_j \bs q_j^{(m)}), \tx{ with }
    q_{jm^\prime}^{m} =
    \begin{cases}
        \delta_j : & m^\prime = m \\
        1-\delta_j : & m^\prime \neq m
    \end{cases}
\end{gather}
where $1/2 \leq \delta_j \leq 1$ and $\nu_j>0$ is a fixed constant, such as $\nu_j=1$. At one extreme, when $\delta_j = 1/2$, we have that $\bs p_j^{(l)} \sim \tx{Dir}(\nu_j/2, \dots, \nu_j/2),$ i.e. the MRs of feature $j$ have no impact on the clustering. Hence, the parameter $\nu_j$ controls the distribution of the null model. On the other extreme, as $\delta_j \to 1$, $\bs p_j^{(l)} \sim \sum_{m=1}^{M_j} \phi_j^{(m)} \delta_{\bs e_m}$, where $\bs e_1, \dots, \bs e_{M_j}$ are the standard basis vectors, meaning that clusters correspond exactly to the categories. In the special case in which $y_{ij}$ is binary, we focus on modeling $0<p_j^{(l)}<1$, or the probability that feature $j$ is equal to $1$, via a mixture of Beta distributions. That is, \eqref{eq:multinomial-CLAMR} translates to 
\begin{equation} \label{eq:binary-CLAMR}
    p_j^{(l)} \sim \phi_j \tx{Beta}(\nu_j \delta_j, \nu_j(1-\delta_j)) + (1-\phi_j )\tx{Beta}(\nu_j (1-\delta_j), \nu_j \delta_j),
\end{equation}
where each component is an informative Beta distribution concentrated at the endpoints of the unit interval. Clearly, when $\delta_j = 1/2$, then $p_j^{(l)} \sim \tx{Beta}(\nu_j/2, \nu_j/2)$, indicating no clustering heterogeny. For both the categorical and binary cases, we construct a Bayesian hypothesis test by focusing on $H_0: \delta_j < 1/2 + \epsilon_j$, where $\epsilon_j >0$ is a small positive constant. Observe that, since we assume independence between features, we can combine \eqref{eq:multinomial-CLAMR} and \eqref{eq:binary-CLAMR} with the Gaussian CLAMR model we discuss in the main article in order to apply our approach to data of mixed types. 

\section{Extension to Non-Diagonal Covariance Matricies}
While we focus on the GMM with diagonal covariance in the main article, the general GMM specification has non-diagonal covariance matricies:
\begin{equation*}
    \bs y_i \sim \sum_{l=1}^L \psi_l \N(\bs \mu^{(l)}, \Sigma^{(l)}),
\end{equation*}
where $\bs \mu^{(l)} = (\mu_j^{(l)})_{j=1}^p$ is a vector of cluster centers and $\Sigma^{(l)}$ is the covariance matrix of the $l$th cluster. This model assumes clusters with different shapes than the diagonal-covariance GMM. In the context of INDITe, the non-diagonal covariance GMM allows for more flexible characterization of subtypes, though there is a computational cost, which we detail in the following paragraph. In practice, one could then set the prior for $\Sigma^{(l)}$ to be independent of that for the centers, e.g. a weakly informative inverse-Wishart prior. This ensures that the Gibbs update of $\Sigma^{(l)}$ can be implemented using the typical inverse-Wishart posterior.

To preserve the marginal structure of the CLAMR prior (i.e., GMMs along each feature), we take the prior for $\bs \mu^{(l)}$ as a mixture over all possible combinations of the MRs. In other words, we mix over the entire profile matrix $\bs s$, leading to the following prior:
\begin{equation} \label{eq:multivariate-CLAMR}
    \bs \mu^{(l)} \sim \sum_{k_1=1}^{K_1} \cdots \sum_{k_p=1}^{K_p} \phi^{(k_1, \dots, k_p)} \N(\bs \xi^{(k_1, \dots, k_p)}, \tx{diag}(\bs \tau^{(k_1, \dots, k_p)})).
\end{equation}
Here, $\bs \xi^{(k_1, \dots, k_p)} = (\xi_j^{(k_j)})_{j=1, \dots, p}$ and $\bs \tau^{(k_1, \dots, k_p)} = (\tau_j^{(k)2})_{j=1, \dots, p}$. If $\phi^{(k_1, \dots, k_p)}=\prod_{j=1}^p \phi_j^{(k_j)}$, then we have the exact same CLAMR prior discussed in the main article (i.e., independent GMMs). Hence, we could use \eqref{eq:multivariate-CLAMR} to jointly update all cluster centers. However, this would involve sampling a vector of $p$ MR labels, which can be difficult to do in practice.

More generally, the joint specification of the CLAMR prior in \eqref{eq:multivariate-CLAMR} could extend our methodology to a wide class of conjugate priors for the cluster centers, such as assuming that the within-MR covariance of $\bs \mu^{(l)}$ is non-diagonal. Ultimately, in comparison to the framework presented in the main paper, computation and hyperparameter choice become more challenging when considering all of the features jointly.

\end{document}